\documentclass
[preprint,review,12pt]{elsarticle}

\usepackage[utf8]{inputenc}
\usepackage{fullpage}
\usepackage{amsmath}
\usepackage{subfigure}
\usepackage{fancybox}
\usepackage{tikz}
\usepackage{graphicx}
\usepackage{amssymb}
\usepackage{fancybox}
\usepackage{mathtools}
\usepackage{amsmath}
\usepackage{amsfonts}
\usepackage{adjustbox}
\usepackage{rotating}
\usepackage{amsthm}
\usepackage{enumitem}
\usepackage{xspace}
\usepackage{comment}
\usepackage{bbm}
\usepackage{mlmodern}
\usepackage[titletoc]{appendix}

\usepackage{lineno}

\usepackage{float} 
\usepackage{subfigure}
\usepackage[]{caption2} 

\usepackage{color}   
\usepackage{hyperref}
\hypersetup{
    colorlinks=true, 
    linktoc=all,     
    linkcolor= Blue,  
}

\begin{document}
\tableofcontents

\begin{frontmatter}


\title{A Novel Nonlinear Non-parametric Correlation Measurement With A Case Study on Surface Roughness in Finish Turning}

\author[Northeastern]{Ming Luo}

\author[Northeastern]{Srinivasan Radhakrishnan}

\author[Northeastern]{Sagar Kamarthi}

\address[Northeastern]{Northeastern University, 360 Huntington Avenue, Boston, Massachusetts, United States}


\begin{abstract}
Estimating the correlation coefficient has been a daunting work with the increasing complexity of dataset's pattern. One of the problems in manufacturing applications consists of the estimation
of a critical process variable during a machining operation from directly measurable process variables.
For example, the prediction of surface roughness of a
workpiece during finish turning processes. In this paper, we did exhaustive study on the existing popular correlation coefficients:  Pearson correlation coefficient, Spearman’s rank correlation coefficient, Kendall’s Tau correlation coefficient, Fechner correlation coefficient, and Nonlinear correlation coefficient. However, no one of them can capture all the nonlinear and linear correlations. So, we represent a universal non-linear non-parametric correlation measurement, $g$-correlation coefficient. Unlike other correlation measurements, $g$-correlation doesn't require assumptions and pick the dominating patterns of the dataset after examining all the major patterns no matter it is linear or nonlinear. 
Results of testing on both linearly correlated and non-linearly correlated dataset and comparison with the introduced
correlation coefficients in literature show that $g$-correlation is robust on all the linearly correlated dataset and outperforms for some non-linearly correlated dataset. Results of the application of different correlation concepts to
surface roughness assessment show that $g$-correlation has a central
role among all standard concepts of correlation.
\end{abstract}


\end{frontmatter}


\newpage
\section{INTRODUCTION} A classical problem in statistics for ordered pairs of measurement $(x_1,y_1), (x_2,y_2),\\ \dots, (x_n,y_n)$ is that is it possible in general from knowing one of the values in an arbitrary pair $\,(x,~y)\,$ of measurements to draw conclusions about the other value in this pair.

When speaking about correlation the statistics literature mainly aims at finding a certain functional relationship (such as a straight line in linear regression) or a monotonic relationship between two numerical variables. In General, correlation analysis is a means of measuring the strength or `closeness' of the relationship between two
variables \cite{fleming1994principles}.

In section \ref{basic} we try to provide general definitions and clarifications
of the terms correlation and predictability for pairs of random variables.
The most important methods for measuring a correlation between
two random variables are also presented and discussed in section \ref{basic}.
Two concepts for nonparametric correlation analysis
and prediction $--$ a seldomly used one and the recently
introduced $g$-correlation $--$ which can
detect correlations which are neither functional nor monotonic relationships
are described, derived and analyzed in sections \ref{section3} and \ref{section4}. In section \ref{section5} it is
shown how the $g$-correlation concept
can be generalized for more than 2 variables. In section \ref{section6} the $g$-correlation
is compared with among all the correlation measures introduced in the paper with linearly and nonlinearly correlated dataset. In section  \ref{section7} the $g$-correlation
is compared with other important correlation measures for a problem
of surface roughness prediction in finish turning and the results are
summarized in section~\ref{section8}.\\
\section{THE CONCEPT OF CORRELATION BETWEEN TWO VARIABLES} \label{basic}

When prediction is a concern in correlation analysis, statistician make the
following distinction \cite{richards1998zen}:
Correlation is a measure of degree of dependence between two dependent random variables.

\noindent\underline{\bf Definition 1} {\sl A dependent random variable $Y$ is
uncorrelated with respect to an independent random variable $X$ if
the range and/or frequency of the possible values for $Y$ is constant
for varying values of $X$. In mathematical terms, this means that, for any two unique values $x_i, ~x_j$ in $X$, $\,F_{Y|X}(y|x_i):=P(Y\leq y|X=x_i)\,$ and
$\,F_{Y|X}(y|x_j):=P(Y\leq y|X=x_j)\,$ are equivalent for all $y$ in $Y$.}

Correlation therefore means that the values of the dependent variable $Y$
do not always behave completely randomly. In fact, their distribution is
influenced by and can be predicted from the corresponding value of $X$.
It is shown with the following Lemma 2 that under some technical
assumptions the correlation defined by Definition 1 is equivalent to the
well-known concept of statistical dependence of two random variables --
one random variable has a certain impact on the other random variable
and vice versa.

\noindent\underline{\bf Lemma 2} {\sl Let $X$ and $Y$ be two random variables
which are either a) both discrete or b) both continuous with a joint density
function $f$. Under this assumption, $X$ and $Y$ are uncorrelated according to
Definition~1 if and only if they are statistically independent.}\\

\begin{proof} To prove it, we will show that the statement that X and Y are uncorrelated is both necessary and sufficient for statement that X and Y are  statistically independent in both case a) and case b).

\noindent {\sl \\Case a:} The discrete case means that $X$ and
 $Y$ take finitely or countably many values $\,x_1,~x_2,~\dots\,~x_j,~\dots\,~x_n$
and $\,y_1,~y_2,~\dots,~y_i,~\dots\, ~y_n$, respectively, which are listed in increasing order.
In this case statistical
independence of $X$ and $Y$ is defined as
\begin{equation}\label{eq1}
    {\rm P_{Y,X}}\{Y=y_i,~X=x_j\}={\rm P_Y}\{Y=y_i\}{\rm P_X}\{X=x_j\}
\end{equation}
for all indices $\,i,j$.\\


\begin{enumerate}
\item Necessary: 
\noindent If $X$ and $Y$ are independent, then the conditional probability
distribution function of $Y$ given $X = x,$ $ F_{Y|X}(y|x),$ is defined as
\begin{equation}
    F_{Y|X}(y|x)={\rm P}\{Y\leq y|X=x\} =\frac{{\rm P}\{Y\leq y,~X=x\}}{{\rm P}\{X=x\}}
\end{equation}\\
for all $x \in \{\,x_1,~x_2,~\dots\,~x_j,~\dots\,~x_n\}$ such that ${\rm P}\{X=x\} > 0$. This can be further written as
  \begin{equation}
    F_{Y|X}(y|x)=\frac{\sum_{y_{k} \leq y}{\rm P}\{Y=y_k,~X=x\}}{{\rm P}\{X=x\}}, y_k \in \{\,y_1,~y_2,~\dots,~y_i,~\dots\, ~y_n\}
  \end{equation}
and using equation (\ref{eq1}) we further get
\begin{equation}
    F_{Y|X}(y|x)=\frac{\sum_{y_k\leq y}{\rm P}\{Y=y_k\}{\rm P}\{X=x\}}{{\rm P}\{X=x\}}=\sum_{y_{k} \leq y}{\rm P}\{Y=y_k\}=P\{Y<y\}.
\end{equation}
This means that $\,F_{Y|X}(y|x)\,$ is always the same function independent
of the value $\,x$, i.e. $X$ and $Y$ are uncorrelated.
\item Sufficient: 
For the opposite direction,
  \begin{enumerate}
      \item we first consider the case where
$X$ takes only a single value $x_1$ with probability 1. Then, clearly,
\begin{equation}
    {\rm P}\{Y=y_i,~X=x_1\}={\rm P}\{Y=y_i\}={\rm P}\{Y=y_i\}{\rm P}\{X=x_1\}
\end{equation}
for every index $i$ range from 1 to $n$, which is just the independence of $X$ and $Y$.
      \item Secondly, assume that
\begin{equation}\label{eq6}
    F_{Y|X}(y|x_k)={\rm P}\{Y\leq y|X=x_k\}={\rm P}\{Y\leq y|X=x_j\}=F_{Y|X}(y|x_j)
    \end{equation}
for every real value $y$ and every pair of different indices $\,k,j \in \{1,2,3,\dots, n\}$.
  \end{enumerate}

Equation (\ref{eq6}) can be reformulated as
\begin{align}
    &\frac{{\rm P}\{Y\leq y,~X=x_k\}}{{\rm P}\{X=x_k\}}=\frac{{\rm P}\{Y\leq y,~X=x_j\}}{{\rm P}\{X=x_j\}}\nonumber\\
\Longleftrightarrow \phantom{aaa} &\frac{\sum_{y_l\leq y}{\rm P}\{Y=y_l,~X=x_k\}}{{\rm P}\{X=x_k\}}=\frac{\sum_{y_l\leq y}{\rm P}\{Y=y_l,~X=x_j\}}{{\rm P}\{X=x_j\}}.
\end{align}
Substituting $y$ by the values $y_i$ and $y_{(i-1)}$ for an arbitrary index $i$
leads to
\begin{equation}\label{eq2}
    \frac{\sum_{y_l\leq y_{(i-1)}}{\rm P}\{Y=y_l,~X=x_k\}}{{\rm P}\{X=x_k\}}=\frac{\sum_{y_l\leq y_{(i-1)}}{
\rm P}\{Y=y_l,~X=x_j\}}{{\rm P}\{X=x_j\}}
\end{equation}
and
\begin{equation}\label{eq3}
\frac{\sum_{y_l\leq y_i}{\rm P}\{Y=y_l,~X=x_k\}}{{\rm P}\{X=x_k\}}=\frac{\sum_{y_l\leq y
_i}{\rm P}\{Y=y_l,~X=x_j\}}{{\rm P}\{X=x_j\}}.
\end{equation}
Subtracting equation (\ref{eq2}) from (\ref{eq3}) gives
\begin{equation}
\frac{{\rm P}\{Y=y_i,~X=x_k\}}{{\rm P}(\{X=x_k\}}=\frac{{\rm P}\{Y=y_i,~X=x_j\}}{{\rm P}\{X=x_j\}}
\end{equation}
, which is equivalent to  
\begin{equation}
{\rm P}\{Y=y_i,~X=x_k\}=\frac{{\rm P}\{Y=y_i,~X=x_j\}}{{\rm P
}\{X=x_j\}} \cdot {\rm P}(\{X=x_k\}.
\end{equation}

In addition, for each index $i$ and consequently
\begin{align}
{\rm P}\{Y=y_i\}&=\sum_{k=1}^{n}{\rm P}\{Y=y_i,~X=x_k\}\\
& = \sum_{k=1}^{n}\frac{{\rm P}\{Y=y_i,~X=x_j\}}{{\rm P
}\{X=x_j\}} \cdot {\rm P}(\{X=x_k\} \nonumber\\
&=\frac{{\rm P}\{Y=y_i,~X=x_j\}}{{\rm P}\{X=x_j\}}\sum_{k=1}^{n}{\rm P}\{X=x_k\}\nonumber\\
&=\frac{{\rm P}\{Y=y_i,~X=x_j\}}{{\rm P}\{X=x_j\}}.\nonumber
\end{align}
This implies that
\begin{equation}
{\rm P}\{Y=y_i,~X=x_j\}={\rm P}\{Y=y_i\}{\rm P}\{X=x_j\}
\end{equation}
for all $i$ and $j$, which means that the random variables $X$ and $Y$
are independent.
\end{enumerate}

\noindent {\sl Case b:} In the continuous case, the statistical independence of $X$ and $Y$
is
equivalent to the relation
\begin{equation}\label{eq14}
    f_{YX}(y,x)=f_Y(y)f_X(x)\quad\mbox{almost everywhere,}
\end{equation}
where $f_Y$ and $f_X$ are the marginal probability density functions of $Y$ and $X$ respectively \cite{pfeifferprobability}. 
\begin{enumerate}
\item Necessary: 
The conditional probability density function of $Y$ given $X = x$ is defined by
\begin{equation}
    f_{Y|X}(y|x):=\frac{f_{YX}(y,~x)}{f_X(x)}
\end{equation}
for every value $x$ with $\,f_X(x)>0\,$ \cite{hogg1997sampling}.
If $X$ and $Y$ are independent, then it follows from equation (\ref{eq14}) that
\begin{equation}
    f_{Y|X}(y|x)=f_Y(y)\quad\mbox{almost everywhere}
\end{equation}
for every possible value $x$ with $\,f_X(x)>0$. Thus the conditional cumulative distribution function of $Y$ given $X=x$  is \[\,F_{Y|X}(y|x):=P(Y\leq y|X=x)=\int_{-\infty}^yf_{Y|X}(u|x)\,\mbox{d}\mu(u)\,\] are all equal if $x$ is varied. Consequently, by Definition~1, $X$ and $Y$ are uncorrelated.
\item Sufficient:
Conversly, if we assume that
\begin{equation}
    F_{Y|X}(y|x_i)=F_{Y|X}(y|x_j)
\end{equation}
for every pair of values $\,x_i \neq x_j\, $ for which $\,F_{Y|X}\,$ can be defined. We get
\begin{equation}
    f_{Y|X}(y|x_i)=F_{Y|X}'(y|x_i)=F_{Y|X}'(y|x_j)=f_{Y|X}(y|x_j)\quad\mbox{almost everywhere}.
\end{equation}
If we fix $x_i$ arbitrarily, then
\begin{align}
    \frac{f_{YX}(y,~x_i)}{f_X(x_i)}&=\frac{f_{YX}(y,~x_j)}{f_X(x_j)}\quad\mbox{almost everywhere}\\\nonumber
    \Longleftrightarrow f_{YX}(y,~x_j)&=\frac{f_{YX}(y,~x_i)}{f_X(x_i)}f_X(x_j)\quad\mbox{almost everywhere}
\end{align}
for every value $x_j$. Consequently,
\begin{align}
   f_Y(y)  & =\int f_{YX}(y,~x_j)\,\mbox{d}\mu(x_j) \\ &=\int\frac{f_{YX}(y,~x_i)}{f_X(x_i)}f_X(x_j)\,\mbox{d}\mu(x_j)\\ & =\frac{f_{YX}(y,~x_i)}{f_X(x_i)}\int f_X(x_j)\,\mbox{d}\mu(x_j)\\ &
   = \frac{f_{YX}(y,~x_i)}{f_X(x_i)}\quad\mbox{almost everywhere},
\end{align}
which is equivalent to equation (\ref{eq14}), which means that $X$ and $Y$ are statistically independent.\\
If, on the other hand, $x_i$ is such that $\,f_X(x_i)=0$, then
\begin{equation}
    f_Y(y)f_X(x_i)=0
\end{equation}
and
\begin{equation}
    0=f_X(x_i)=\int f_{YX}(y,~x_i)\,\mbox{d}\mu(y)
\end{equation}
implies
\begin{equation}
    f_{YX}(y,~x_i)=0
\end{equation}
for almost every $y$ because the integrand is nonnegative.\hfill\\
\end{enumerate}

\end{proof}
\noindent Figure \ref{fig:subfig:a} shows a case of no correlation  between the variables $X$ and $Y$. One of the reasons for the study of correlation between two variables is to seek a functional relationship between two random variables (See \cite{drake1996assessment} and Figure \ref{fig:subfig:b} for examples.). However, when it is not possible to establish a functional relationship between $X$ and $Y$ (see Figure \ref{fig:subfig:c} for example), then measuring correlation has not been sufficiently dealt with in the past. Such a situation occurs in the prediction of a surface roughness in a turning operation. The development and application of a correlation concept for such a scenario is the objective of this paper.
Existing correlation measures are briefly introduced in next section.
Let $\,x_1,~x_2,~\dots,~x_n\,$ and
$\,y_1,~y_2,~\dots,~y_n\,$ be samples of two random variables $X$ and $Y$,
respectively.

\subsection{Pearson Correlation Coefficient} 
The standard Pearson correlation coefficient  \cite{rohatgi1976introduction}
\begin{equation}
    r:=\frac{\sum_{i=1}^n(x_i-\bar x)(y_i-\bar y)}{\sqrt{\sum_{i=1}^n(x_i-\bar x)^2\sum_{i=1}^n(x_i-\bar x)^2}}
\end{equation}
of Pearson \cite{rohatgi1976introduction}, where $\bar x$ and $\bar y$ are the sample means
respectively. The range of correlation coefficient $r$ is [-1,~1]. The closer $r$ is to 1 or -1, the stronger the correlation between the two random variables $X$ and $Y$. The degree of the linear dependency between $X$ and $Y$ can be measured through $|r|$: $\,|r|=1\,$ if and only if the points $(x_i,~y_i)$ describe a straight line in $\mathbb{R}^2$ which is neither horizontal nor vertical. 

For the random variable $X$ and $Y$ in Fig. \ref{fig:subfig:a}, $|r| = -0.136$. Since pearson correlation only detect the linear dependency, so if $X$ and $Y$ are nonlinearly correlated, the pearson correlation is still close to 0. For example, the $r$ for Fig. \ref{fig:subfig:b} is 0.015. 


\subsection{Spearman's Rank Correlation Coefficient} 
We present the Spearman's rank correlation which is a nonparametric correlation coefficient for two numerical variables denoted by
 $\rho$, which ranges from -1 to 1 \cite{rohatgi1976introduction}. And the more the $|\rho|$ is closer to 0, the less association between $X$ and $Y$.   First, sort $\,x_1,~x_2,~\dots,~x_n\,$ and $\,y_1,~y_2,~\dots,~y_n\,$  in an
ascending order. Next, complete a sequence $\,\alpha_1,~\alpha_2,~\dots,~\alpha_n\,$
in which $\alpha_i$ is the position of the corresponding
element $x_i$ in the sorted sequence i.e. $\alpha_i$ = 1 if $x_i$ is the smallest ,  $\alpha_i$ = 2 if
$x_i$ is the second smallest and so on. In a similar fashion, create a
sequence $\,\beta_1,~\beta_2,~\dots,~\beta_n\,$ of ranks corresponding to the
sequence $\,y_1,~y_2,~\dots,~y_n\,$  defined as
\begin{equation}
    \rho=1-\frac{6\sum_{i=1}^n(\alpha_i-\beta_i)^2}{n(n^2-1)}.
\end{equation}
The Spearman's rank correlation coeffecient $|\rho|$ determines the degree to which a monotonic relationship exists between
the two variables $X$ and $Y$.

Fig. \ref{fig:subfig:a} has no apparent monotonic relationship, so its $ |\rho|$ = 0.108. Although Fig. \ref{fig:subfig:b} has clear nonlinear relationship, but due to the limitation of Spearman's, the $|\rho|$  is also very small, which is 0.034. Fig. \ref{fig:subfig:c} has a rough monotonic increasing relationship, the $|\rho|$ = 0.889, which is close to 1. 


\subsection{Kendall's Tau} 
The Kendall nonparametric measure $\tau$ of correlation
between $X$ and $Y$ \cite{hollander1999nonparametric} is defined as
\begin{equation}
    \tau:=\frac{2\sum_{i=1}^{n-1}\sum_{j=i+1}^n\Delta_{ij}}{n(n-1)} ,
\end{equation}
where\\
\begin{equation}
    \Delta_{ij}:=\left\{
    \begin{array}{cc}
    1 & \text{if}\phantom{a} (x_j-x_i)(y_j-y_i)>0\cr
0 & \text{if}\phantom{a} (x_j-x_i)(y_j-y_i)=0\cr
-1 & \text{if}\phantom{a} (x_j-x_i)(y_j-y_i)<0
    \end{array}\right.
\end{equation}
If all values in both of the sequences $\,x_1,~x_2,~\dots,~x_n\,$ and $\,y_1,~y_2,~\dots,~y_n\,$ are different, then Kendall's tau for these two sequences is equal to another correlation measure, which is called Goodman and Kruskal's gamma \cite{freeman1965elementary}. The range of $\tau$ is also between [-1, 1], which is similar as Pearson correlation coefficient and Spearman's rank correlation coefficient. If $|\tau|$ is close to 0, it means very small ordinal association between $X$ and $Y$ is found. Inversely, when $|\tau|$ is close to 1, $X$ and $Y$ have strong ordinal association. 

Fig. \ref{fig:subfig:a} doesn't have an obvious ordinal association between $X$ and $Y$, so its $|\tau|$ is close to 0, $|\tau| = 0.070$. Fig. \ref{fig:subfig:b} has ordinal association between $X$ and $Y$ only for the first half observations and an inverse ordinal association for the second half observations, so its $|\tau|$ is also close to 0, $|\tau| = 0.033$. However,  Fig. \ref{fig:subfig:c} has an obvious ordinal association between $X$ and $Y$, so its $|\tau|$ is relatively large, which is 0.707.

\subsection{Nonlinear Correlation Coefficient } \label{NCC}

For two variables $X, Y$, we can estimate its correlation  by calculate their mutual information after sorting and grouping their values \cite{5382400}. Given the discrete variables $X = \{x_1,x_2, ..., x_{n-1}, x_{n} \}, Y = \{y_1,y_2, ..., y_{n-1}, y_{n} \}$:  

\noindent \underline{\bf Step 1}\, Sort $\{x_i\}, \{y_i\}, i = 1,2,..,n$ in ascending order. Then, we use $X^{s}, Y^{s}$ to represent the ordered $X, Y$. $X^{s} = \{x_{(1)},x_{(2)}, ..., x_{(n-1)}, x_{(n)} \}, Y^{s} = \{y_{(1)},y_{(2)}, ..., y_{(n-1)}, y_{(n)} \}$, and $ x_{(1)}\leq x_{(2)}\leq ...\leq x_{(n-1)}\leq x_{(n)},y_{(1)}\leq y_{(2)}\leq ...\leq y_{(n-1)}\leq y_{(n)} $

\bigskip

\noindent \underline{\bf Step 2}\, Split $X^{s}, Y^{s}$ into $b$ ranks , and each rank contains $\frac{n}{b}$ observations.

\bigskip

\noindent \underline{\bf Step 3}\, For the pair $\{(x_{(i)},y_{(i)})\}, i=1,2,..n,$ we split them into $b^2$ regions.

\bigskip

\noindent \underline{\bf Step 4}\,  Calculate noncorrelation coefficient between $X$ and $Y$ with the mutual information based on the ranks

The nonlinear correlation coefficient (NCC) is as below: 
\[ NCC(X;Y) = H(X^s) + H(Y^s) -H(X^s,Y^s)\], $H(X^s), H(Y^s)$ are the entropy for $X^s ,Y^s$, which are calculated based on $b$ ranks, and $H(X^s,Y^s)$ is the joint entropy,which are calculated based on $b^2$ regions.

\[H(X^s) =-\sum_{i=1} ^b \frac{n_i}{n} \log_b \frac{n_i}{n}= -\sum_{i=1} ^b \frac{n/b}{n} \log_b \frac{n/b}{n} = 1,\] 
\[H(Y^s) =-\sum_{j=1} ^b \frac{n_j}{n} \log_b \frac{n_j}{n}= -\sum_{j=1} ^b \frac{n/b}{n} \log_b \frac{n/b}{n} = 1,\]
and 
\[H(X^s,Y^s) = -\sum_{i=1} ^b \sum_{j=1} ^b \frac{n_{ij}}{n}\log_b \frac{n_{ij}}{n}, \]
$b$ is the base for logarithm, $n_i, n_j$ are the number of observations in $i$-th,$j$-th rank, $n_{ij}$ is the number of observations in $(i,j)$ ragion. 

Thus, \[NCC(X;Y) = 2+\sum_{i=1} ^b \sum_{j=1} ^b \frac{n_{ij}}{n}\log_b \frac{n_{ij}}{n}\].

The range for this nonlinear correlation coefficient is $[0,1]$. And 1 means that a strong nonlinear relationship is being detected and 0 means no nonlinear relationship is being found.  

\begin{figure*}
  \centering
  \subfigure[]{
    \includegraphics[scale=0.35]{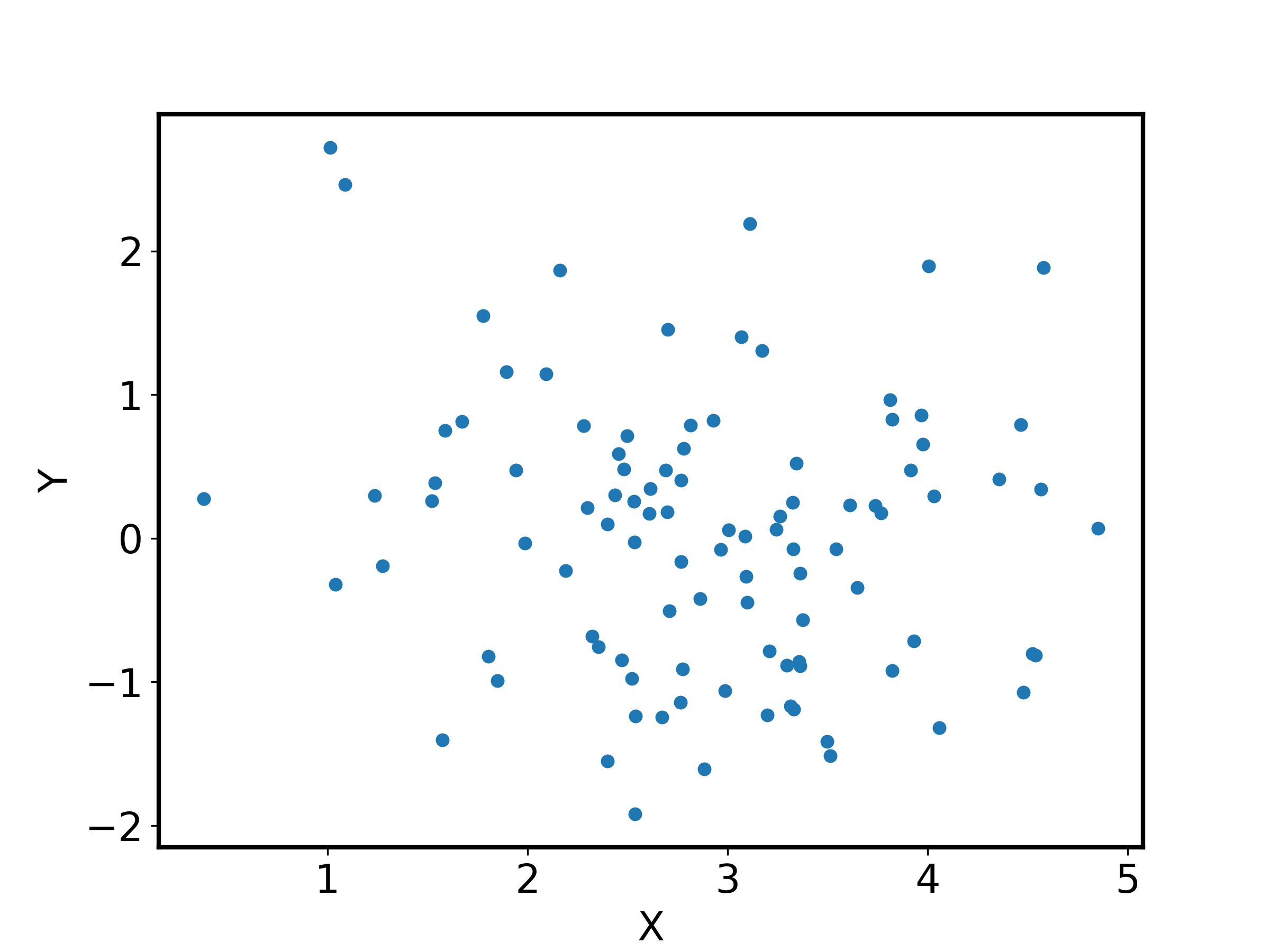}\label{fig:subfig:a}}\\
  \subfigure[]{
    \includegraphics[scale=0.35]{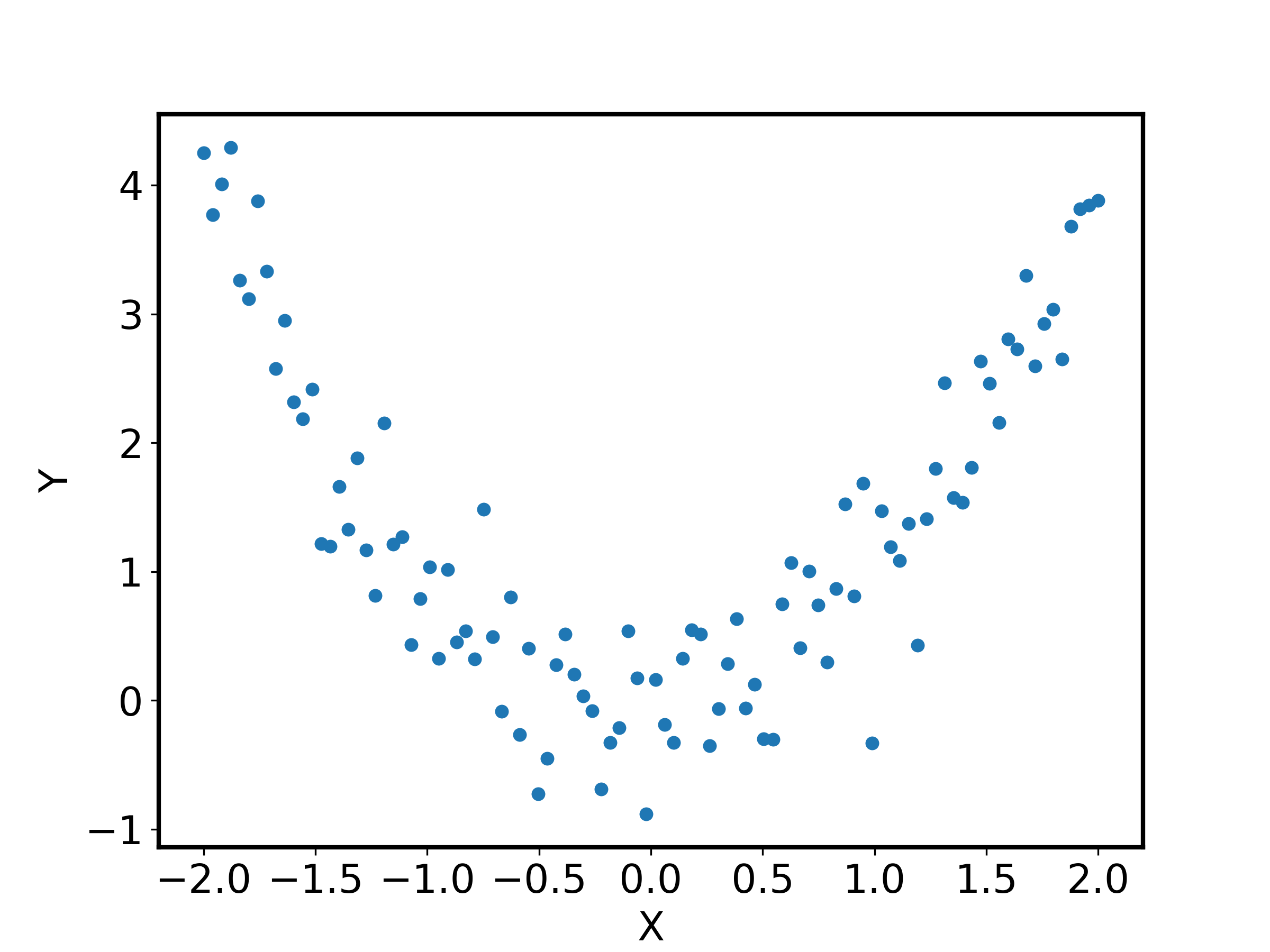}\label{fig:subfig:b}}
    \subfigure[]{
    \includegraphics[scale=0.35]{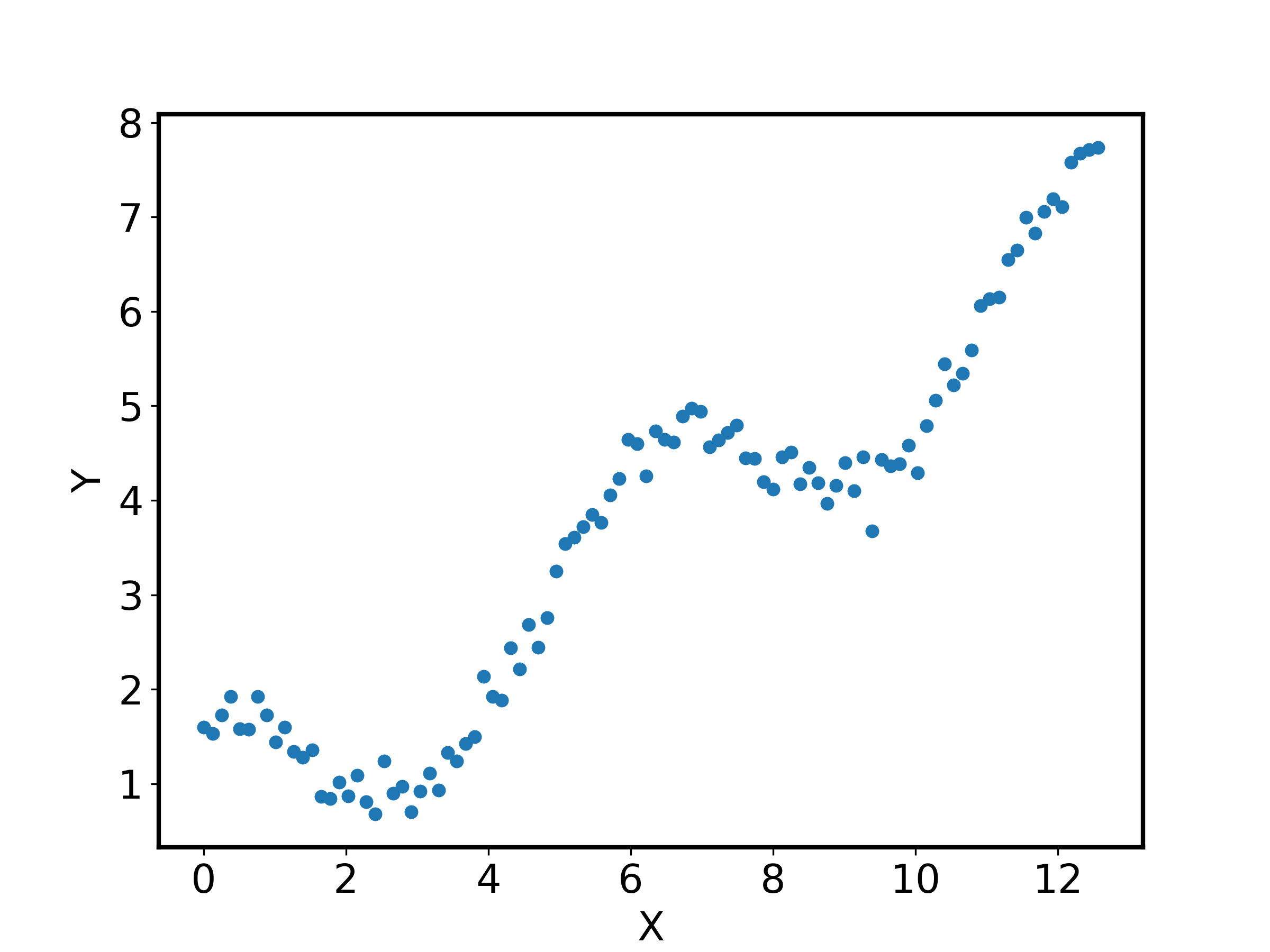}
    
    \label{fig:subfig:c}}
    
    \caption{{(a) Uncorrelated meaurements, (b) curvilinear correlation, (c) correlation for unknown and coars shape a correlation which seems to allow only a coarse estimate of the
dependent variable Y}}
  
\end{figure*}

\begin{figure*}
    \centering
    \includegraphics[scale=0.8]{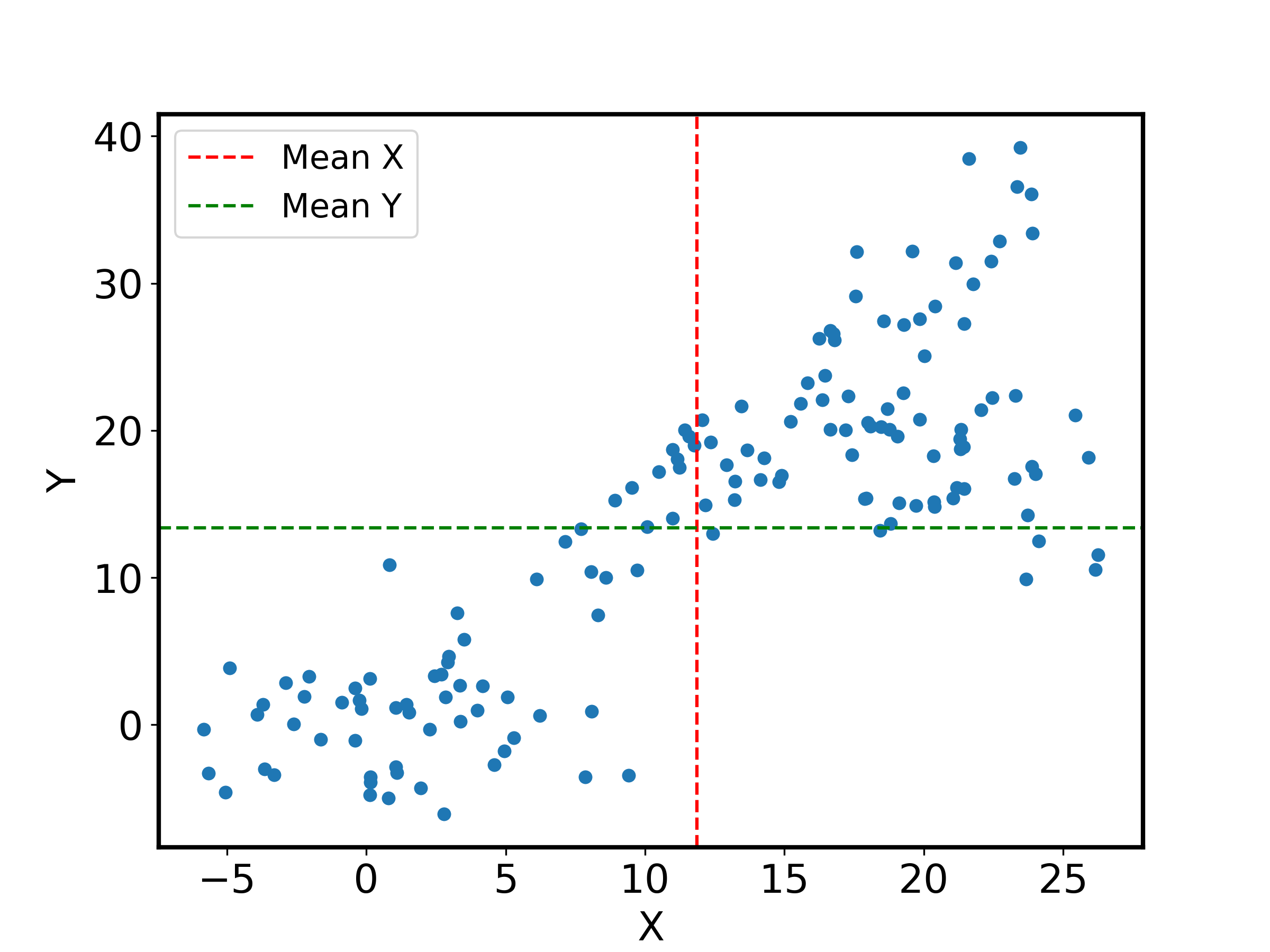} 
    \caption{Demonstration of Fechner correlation coefficient ($\kappa =$0.907). The data points are separated into 4 areas by vertical line $x = \bar{x}$ and horizontal line $y = \bar{y}$.}
\label{fig:subfig:demo-fech}

\end{figure*}

We choose $b=10$ for calculating the nonlinear correlation coefficient. If $X$ and $Y$ are not related, the $NCC$ is small (see example Fig. \ref{fig:subfig:a}, whose $NCC$ is 0.239). $NCC$ for Fig. \ref{fig:subfig:b} and Fig. \ref{fig:subfig:c} are 0.433 and 0.716, which are larger since these two figures has more clear pattern than Fig. \ref{fig:subfig:a}. In addition, for Fig. \ref{fig:subfig:demo-fech}, $NCC=0.370$. And for Fig. \ref{fig:fech_flaw}, $NCC=1$. We can see from the results that when $X$ and $Y$ has a more clear nonlinear relationship, its $NCC$ is relatively larger.  

\section{FECHNER CORRELATION COEFFICIENT}\label{section3}

In this section Fechner correlation coefficient \cite{Latyshev_Koldanov_1970a}
is
reviewed which is not as widely known in the literature.
\subsection{Definition and Interpretation}

The Fechner correlation coefficient is defined as
\begin{equation}
    \kappa:=\frac{1}{n}\sum_{i=1}^n\mbox{sign}(x_i-\bar x)\mbox{sign}(y_i-\bar y),
\end{equation}
where $\bar x$ and $\bar y$ are the sample means of the
sequences $\,(x_1,~x_2,~\dots,~x_n)\,$ and $\,(y_1,~y_2,~\dots,~y_n)$,
respectively, and
\begin{equation}
    \mbox{sign}(u):=\left\{
    \begin{array}{cc}
1 & \text{if}\phantom{a}  u\geq 0\cr
-1 & \text{if}\phantom{a} u<0
    \end{array}\right.
\end{equation}
is the sign function. Fechner correlation coefficient $\kappa$ is calculated using the following scheme:

\noindent \underline{\bf Step 1}\, The sequence
$\,((x_1,~y_1),~(x_2,~y_2),~\dots,~(x_n,~y_n))\,$ is sorted
in an based on $x_i$. Let $i_0$ denote the largest index $i$ with
$\,x_i<\bar x\, , 1\leq i\leq n$ .

\bigskip

\noindent \underline{\bf Step 2}\, The sequence from Step 1 is converted to a binary sequence $\,b=(b_1,~b_2,~\dots,~b_m) \\ (m\le n)\,$ by replacing an element $\,(x_i,~y_i)\,$ by 0, if $\,y_i<\bar y\,$ and by 1, if $\,y_i \geq \bar y$.

\bigskip

\noindent \underline{\bf Step 3}\, The Fechner correlation coefficient is then calculated as
\begin{equation}
    \kappa=\frac{1}{n}\left(\sum_{i=1}^{i_0}(1-2b_i)+\sum_{i=i_0+1}^n(2b_i-1)\right).
\end{equation}

\bigskip

Note that $\kappa$ = 1 if the sequence b has the form
$b=(0,~0,~\dots,~0,~1, ~1,~\dots,~1)$
with the jump from 0 to 1 occuring at the indices
$\,i_0+1$. On the other hand, $\kappa$ is equal to $-1$
if the sequence $b$ has the form
$b=(1,~1,~\dots,~1,~0,~0,~\dots,~0)$
with the jump from 1 to 0 occuring again at the indices
$\,i_0+1$. Figure \ref{fig:subfig:demo-fech}, $\,\kappa= 0.907$.
For Figure \ref{fig:subfig:a}, $\,\kappa=-0.020$ and for Figure \ref{fig:subfig:c}, the $\kappa$
= 0.580.

As in the case for a straight line associated with the standard
Pearson correlation coefficient, $\kappa$ is related to a prediction
model. For $\,|\kappa|\approx 1\,$ it provides a classification
scheme for classifying the values $y$ given the values $x$
\begin{align}\label{clpr}
    & y<\bar y\quad\text{if}\quad (x-\bar x)\mbox{sign}(\kappa)<0,\nonumber\\
& y=\bar y\quad\text{if}\quad x=\bar x,\\
& y>\bar y\quad\text{if}\quad (x-\bar x)\mbox{sign}(\kappa)>0\nonumber
\end{align}


\subsection{Properties}

One of the drawbacks of the Fechner correlation
coefficient is that it does not
provide any insight of the shape of the data
$\,\{(x_i,~y_i):i~=~1,~2,~\dots,~n\}$. However demonstrated
in Figure \ref{fig:subfig:c} and Figure \ref{fig:subfig:demo-fech}, due to the information reduction in Step 1 and 2, $\kappa$ permits the detection of correlations
even when accurate predictions of $Y$ are not possible. This can be a big
advantage of the Fechner correlation coefficient, at least in certain cases.

Assume that data points $\,(x_i,~y_i)\,$ lie on a straight line of
the form $y=ax+b$.

If $\,a=0$, then
\begin{equation}
    \kappa=\frac{1}{n}\sum_{i=1}^n\mbox{sign}(x_i-\bar x)\simeq \frac{\frac{n}{2} +( - \frac{n}{2})}{n} = 0.
\end{equation}
For $\,a\ne 0\,, $ for $\,i~=~1,~2,~\dots,~n\,$ it follows that
\begin{equation}
    \bar y=\frac{1}{n}\sum_{i=1}^ny_i=\frac{1}{n}\Bigl(a\sum_{i=1}^nx_i+nb\Bigr)=a\bar x+b
\end{equation}
consequently the Fechner correlation coefficient becomes
\begin{equation}
  \kappa=\frac{1}{n}\sum_{i=1}^n\mbox{sign}(x_i-\bar x)\mbox{sign}[a(x_i-\bar x)]= \frac{1}{n}\sum_{i=1}^n a (\mbox{sign}(x_i-\bar x))^2= \mbox{sign}\,a.
\end{equation}
That is $\,|\kappa|=1\,$ if the points
$\,(x_i,~y_i)\phantom{a} \text{for}\phantom{a}  i~=~1,~2,~\dots,~n\,$ lie on a straight line $--$ a property shared with the correlation coefficients.

However, often data can not be a sample from a strictly monotonically
increasing function for which $\kappa$ indicates that they are uncorrelated
with a small negative correlation. See Figure \ref{fig:fech_flaw}.
In the next section,the Fechner correlation coefficient is improved
to handle such cases.


\section{$g$-CORRELATION}\label{section4}

As discussed earlier, the Fechner correlation
coefficient $\kappa$ need not detect monotonic relationships between $X$ and $Y$ as
opposed to the correlation measures presented in the subsections 3.B
and 3.C. It is shown that the Fechner correlation coefficient can be
improved by splitting the data points by a vertical and a horizontal
line in a more sensible way instead of arbitrarily dividing the data
into 4 classes based on the lines of $\,x=\bar x\,$ and $\,y=\bar y$


\subsection{Definition} \label{g-def}
As a first step consider the line $\,y=\tilde{y}\,$, where
$\tilde{y}$ is the median of $Y$,
to divide the space of measurements into the following two classes
\begin{equation}\label{defmu}
    C_1:=\{(x,~y)\in\mathbb{R}^2~|~y> \tilde{y}\}\quad\mbox{and}\quad C_2:=\{(x,~y)\in\mathbb{R}^2~|~y< \tilde{y}\}.
\end{equation}

Assume that the distribution function of $Y$ is
continuous and strictly monotonically increasing or decreasing with respect to $X$. 
\noindent {\sl \\Case 1:} The number of observations of a given dataset,$~n$, is even.

Due to the property  of $\tilde{y}$, we have: 
\begin{equation}
 \begin{aligned}
     {\mbox P}(C_1) &= \frac{1}{n}\sum_{i=1}^n[\mathbbm{1}\{y_i > \tilde{y}\}] = \frac{\frac{n}{2}}{n}=\frac{1}{2}\,, 1\leq i \leq n.\\
     {\mbox P}(C_2) &= 1-{\mbox P}(C_1) = \frac{1}{2}.
 \end{aligned}
\end{equation}

And, 
 \begin{equation}
\mathbbm{1}\{y>\tilde{y}\}:=
  \begin{cases}1~&{\text{ if }}~ y> \tilde{y} ~\\0~&{\text{ if }}~y < \tilde{y} ~\end{cases}
\end{equation}.

\noindent {\sl \\Case 2:} The number of observations of a given dataset,$~n$, is odd.

The above assumption means that there is only one data point $(x_m,y_m), ~m\in [1, n]$ such that $y_m = \tilde{y}$. 

Assume $n$ is large and due to the property  of $\tilde{y}$, we have: 
\begin{equation}
\begin{aligned}
    {\mbox P}(C_1) &= \frac{1}{n}\sum_{i=1}^n[\mathbbm{1}\{y_i > \tilde{y}\}] = \frac{\frac{n-1}{2}}{n}\simeq \frac{1}{2}\,,\\
{\mbox P}(C_2) &= 1-{\mbox P}(C_1)-\frac{1}{n} = \frac{\frac{n-1}{2}}{n}\simeq \frac{1}{2}\,.
\end{aligned}
\end{equation}

Thus, each of the classes $C_1$ and $C_2$ contain about half of the observations in the
data set, leading to an
optimal separation.

Instead of choosing the fixed line $\,x=\bar x\,$ for segmenting the plane that is formed by the dataset into 4 classes $C_1^{+},~  C_1^{-},~  C_2^{+},~  C_2^{-}$, and 

\begin{equation}\label{4classes}
 \begin{aligned} 
C_1^{+}&:=\{(x,~y)\in\mathbb{R}^2~|~ x>c,~ y> \tilde{y}\}\quad \\ 
C_1^{-}&:=\{(x,~y)\in\mathbb{R}^2~|~ x\leq c,~ y> \tilde{y}\}\quad\\
C_2^{+}&:=\{(x,~y)\in\mathbb{R}^2~|~ x>c,~ y < \tilde{y}\}\quad \\ 
C_2^{-}&:=\{(x,~y)\in\mathbb{R}^2~|~ x\leq c,~ y < \tilde{y}\},\quad
 \end{aligned} 
\end{equation}

we will use the optimum line:

\bigskip

\noindent\underline{\bf Definition 3} {\sl Two random variable $X$ and $Y$ are said to be
correlated
if there exists $\,c\in\mathbb{R}\,$ such that the criterion
\begin{equation}\label{crit}
    x = c
\end{equation}


assigns realizations $\,(x_i,~y_i)\,$ of
$\,(X,~Y)\,$ to class $C_1^{+}$ or $C_1^{+}$ if they are classified as $C_1$ based on equation (\ref{defmu}), or class $C_2^{+}$ or $C_2^{+}$ if they are classified as $C_2$. And the g-correlation is 
\begin{equation}
   \operatorname*{argmax}_c g(c) := \max\{{\mbox P}(~C_1^{+})+{\mbox P}(~C_2^{-}),~{\mbox P}(~C_1^{-})+{\mbox P}(~C_2^{+})\}. 
\end{equation}

The supremum (the largest upper bound) of all such classification probabilities obtained via the
equation (\ref{crit}) for different $c$ is called the
$g$-correlation coefficient of $X$ and $Y$.}




\subsection{Properties}

\

\noindent\underline{\bf Lemma 4} {\sl The range of g-correlation coefficient of $X$ and $Y$ is [0.5, 1].}

\begin{proof}

For any given $c$,
\begin{equation}
g(c) = \max\{{\mbox P}(~C_1^{+})+{\mbox P}(~C_2^{-}),~{\mbox P}(~C_1^{-})+{\mbox P}(~C_2^{+})\}, 
\end{equation}
the restrictions for the above equation are , 
\begin{equation}\label{restriction}
 \begin{aligned}
P(C_1^{+})& + P(C_1^{-})+P(C_2^{+}) + P(C_2^{-}) = 1\\
P(C_1^{+})& + P(C_1^{-}) =  P(C_1) = 0.5 \\ 
P(C_2^{+})& + P(C_2^{-}) =P(C_2)=  0.5 \\
P(C_1^{+}) &= m P(C_2^{+})\\
P(C_1^{-}) &= n P(C_2^{-}) \\
 \end{aligned}
\end{equation}

In addition, $m,n >0$, this is because when $c$ changes, the number of observations on the same side of vertical line $x=c$ will increase or decrease at the same time. 

\begin{equation}\label{g(c)}
\begin{aligned} 
g(c) &= \max\{{\mbox P}(~C_1^{+})+{\mbox P}(~C_2^{-}),~{\mbox P}(~C_1^{-})+{\mbox P}(~C_2^{+})\}\\
& = \max\{{\mbox P}(~C_1^{+})+{\mbox P}(~C_2^{-}),~ 1-({\mbox P}(~C_1^{+})+{\mbox P}(~C_2^{-}))\}\\
& = \max\{m \cdot {\mbox P}(~C_2^{+})+{\mbox P}(~C_2^{-}),~ 1-({\mbox P}(~C_1^{+})+{\mbox P}(~C_2^{-}))\}\\
& = \max\{(m-1) \cdot {\mbox P}(~C_2^{+})+{\mbox P}(~C_2^{+})+{\mbox P}(~C_2^{-}),~ 1-({\mbox P}(~C_1^{+})+{\mbox P}(~C_2^{-}))\}\\
& = \max\{(m-1) \cdot {\mbox P}(~C_2^{+})+0.5,~1-((m-1) \cdot {\mbox P}(~C_2^{+})+0.5)\}\\
& = \max\{(m-1) \cdot {\mbox P}(~C_2^{+})+0.5,~(1-m) \cdot {\mbox P}(~C_2^{+})+0.5\}\\
\end{aligned}
\end{equation}

In addition, since ${\mbox P}(~C_2^{+})$ is a probability, thus ${\mbox P}(~C_2^{+}) \ge 0$. 

When $m-1\ge 0$,
\begin{equation}
    \begin{aligned}
        (m-1) \cdot {\mbox P}(~C_2^{+})+0.5 \ge 0.5,\\ 
        (1-m) \cdot {\mbox P}(~C_2^{+})+0.5 \le 0.5, 
    \end{aligned}
\end{equation}
and the equation (\ref{g(c)}) equals to: \begin{equation}\label{g(c)min1}
    g(c) = (m-1) \cdot {\mbox P}(~C_2^{+})+0.5 \ge 0.5.
\end{equation}

Similarly, when $(m-1) <0$ and $m>0$,
the equation (\ref{g(c)}) equals to: \begin{equation}\label{g(c)min2}
    g(c) = (1-m) \cdot {\mbox P}(~C_2^{+})+0.5 \ge 0.5.
\end{equation}

Therefore, $g(c) \ge 0.5$ all the time. 

And according to equation (\ref{restriction}), \[{\mbox P}(~C_1^{+})=0\], and \[{\mbox P}(~C_1^{-})={\mbox P}(~C_2^{-})=0.5.\]

Moreover, from equation (\ref{restriction}), we can get that 
\begin{equation}
    0 \le {\mbox P}(~C_1^{+})+{\mbox P}(~C_2^{-}) \le 1, 
\end{equation}
thus, the maximum of equation (\ref{g(c)}) equals to 1 when ${\mbox P}(~C_1^{+})+{\mbox P}(~C_2^{-}) = 1$ and ${\mbox P}(~C_1^{-})+{\mbox P}(~C_2^{+}) = 0,$ or
${\mbox P}(~C_1^{+})+{\mbox P}(~C_2^{-}) = 0$ and ${\mbox P}(~C_1^{-})+{\mbox P}(~C_2^{+}) = 1.$

The range for $g(c)$ is $[0.5,~1]$, thus $g$-correlation coefficient of $X$ and $Y$ ranges from 0.5 to 1. 

And when $g$-correlation is 0.5, it means the $X$ and $Y$ are not correlated. When $g$-correlation is 1, $X$ and $Y$ are perfectly correlated.
\end{proof}


When we define $g$-correlation in section \ref{g-def}, we assume that the distribution function of $Y$ is continuous and strictly monotonically
increasing or decreasing with respect to $X$. But, if we loose the assumption, $g$-correlation still works. However, we need to remove all the data points $(x_i, y_i)$, who share the same trait: $y_i = \tilde{y}$, after finding the $\tilde{y}$ with original dataset. Then use the new modified dataset to calculate $g$-correlation. 

Note that if the new modified dataset has 0 data points, which means the $Y$ is constant, we don't need to calculate the $g$-correlation since $X$ and $Y$ are uncorrelated for sure based on definition 1 in section \ref{basic}. Similarly, if $X$ is constant and $Y$ varies, $X$ and $Y$ are uncorrelated as well.

\begin{figure*}
    \centering
    \includegraphics[scale=0.8]{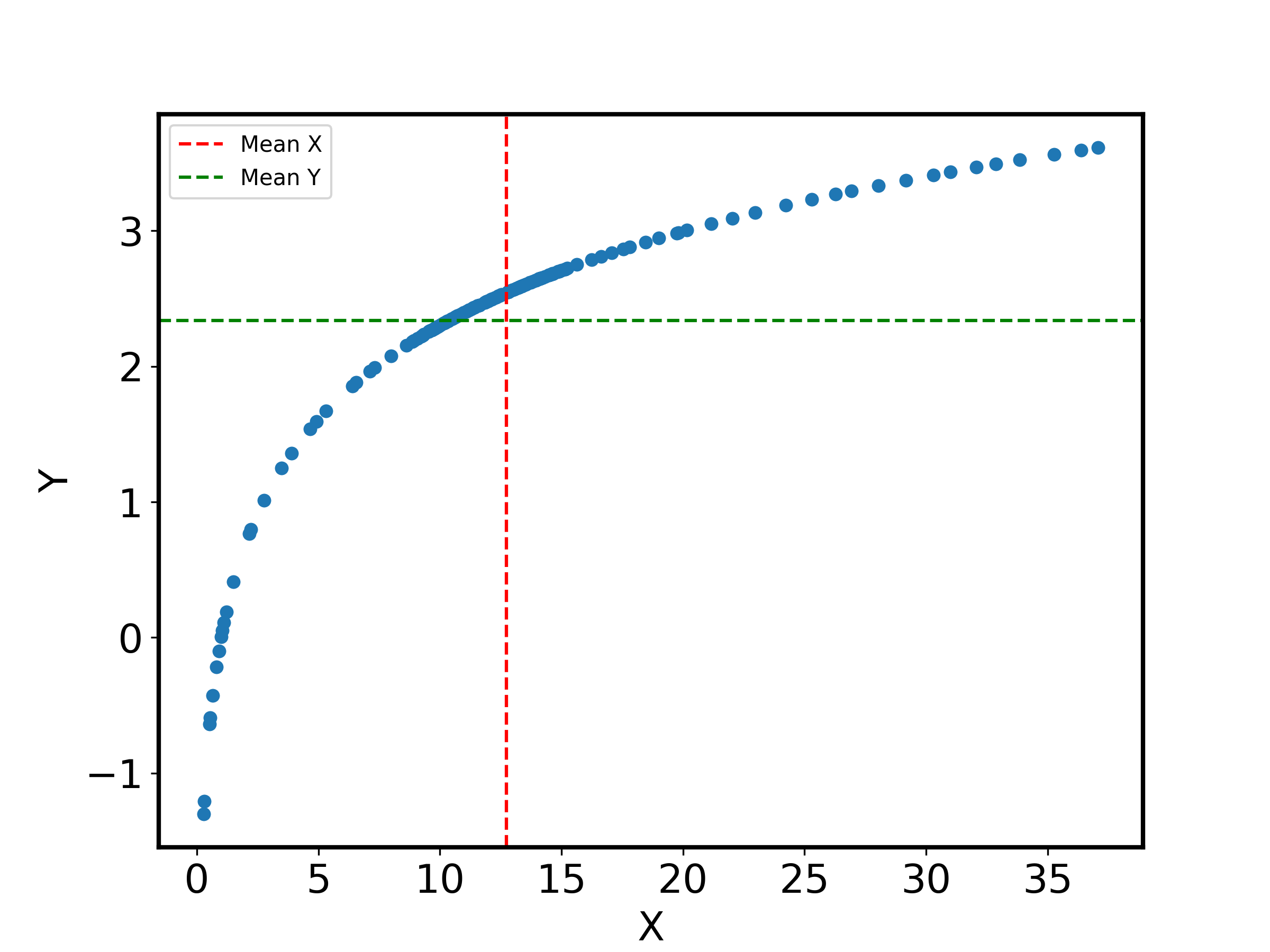}
    
    \caption{Data lie on a strictly monotonically increasing function but which are considered to be uncorrelated by the Fechner correlation coefficient with $\kappa = 0.016$}
    \label{fig:fech_flaw}
\end{figure*}

The $g$-correlation coefficient $\omega$ is in general, not symmetric,
as shown in Figure \ref{fig:fech_flaw}. In that respect $\omega$ differs
from the rest of the correlation coefficients described earlier.
With respect to a $g$-correlation
of $X$ and $Y$ in Figure \ref{fig:fech_flaw}, the set
\begin{equation}
    \{(x_0,~y_0):x_0> c,~y_0>\tilde{y}\},
\end{equation}
where $\,(x_0,~y_0)\,$ are realizations of the random vector
$\,(X,~Y)$, contains 50\% on the average and the set
\begin{equation}
    \{(x_0,~y_0):x_0\le c,~y_0<\tilde{y}\}
\end{equation}
contains 25\% of all measurements on the average. Note that $c$ is optimal because moving the
line $\,x=c\,$ to the left would just decrease the probability

\begin{equation}
    {\rm P}(X \le c,~Y<\tilde{y})
\end{equation}
and moving the line $\,x=c\,$ to the right would just decrease the probability
\begin{equation}
    {\rm P}(X > c,~Y > \tilde{y}).
\end{equation}

The following lemma establishes the main distinction between $\omega$
and the Fechner correlation coefficient:

\bigskip

\noindent\underline{\bf Lemma 6} {\sl Assume $\,Y=f(X)\,$ for a strictly
monotonic continuous function $f(\cdot)$ and the mean of $Y$ is $\tilde{y}$, then $g$-correlated between $X$ and $Y$ can be obtained when $c = f^{-1}(\tilde{y})$, which is 1.}

\bigskip

\begin{proof}

Without loss of generality let $f(.)$ be a
strictly monotonically increasing function. Suppose that $\epsilon >0$, define $\alpha_1$ and
$\alpha_2$ by
\begin{equation}
    \alpha_1:=\sup\{x\in\mathbb{R}:f(x)<\tilde{y}\} +\epsilon\quad\mbox{and}\quad\alpha_2:=\inf\{x\in\mathbb{R}:f(x)>\tilde{y} \}- \epsilon.
\end{equation}
Since $\sup\{x\in\mathbb{R}:f(x)<\tilde{y}\}$ is the largest value of $x$ such that $f(x)<\tilde{y}$, thus, \begin{equation}\label{alpha1}
    \begin{aligned}
        f(x)\ge\tilde{y}, \text{~when~} x \ge \alpha_1 \quad\mbox{and}\quad
    f(x)< \tilde{y}, \text{~when~} x < \alpha_1.
    \end{aligned}
\end{equation} 

Similarly, $\inf\{x\in\mathbb{R}:f(x)>\tilde{y}\}$ is the smallest value of $x$ such that $f(x)>\tilde{y}$, so,
\begin{equation}\label{alpha2}
    f(x)> \tilde{y}, \text{~when~} x > \alpha_2\quad\mbox{and}\quad
    f(x)\le \tilde{y}, \text{~when~} x \le \alpha_2.
\end{equation} 
In addition, if $\,\alpha_1>\alpha_2$, then for every  $\,\xi \in [\alpha_2,~\alpha_1]\,$, according to equation(\ref{alpha1}) and equation(\ref{alpha2}), $f(\xi)>\tilde{y}$ and $f(\xi)<\tilde{y}$, which is a contradiction.

Similarly, if $\,\alpha_1 \le \alpha_2$, then for every $\xi$~ such that $\,\alpha_1 \le \xi \le \alpha_2\,$, we have $\tilde{y} \le f(\xi) \le \tilde{y}$, \text{~which is~}$f(\xi) = \tilde{y}$. 

However, since $\,Y=f(X)\,$ for a strictly
monotonic function $f(\cdot)$, so $x$ and $f(x)$ are one-to-one relationship, and there is a unique $x$ to get the median $\tilde{y}$. Thus, 
\begin{equation}\label{a1=a2}
    x=\alpha_1 = \alpha_2
    \text{~and~}
    f(\alpha_1) =f(\alpha_2) = \tilde{y}
\end{equation}

Then, we will show that the $g$-correlation between $X$ and $Y$ is 1 when we set $c=f^{-1}(\tilde{y})$, which is also $\alpha_1$ . 

Let's split the dataset into 4 classes based on equation (\ref{4classes}) with two lines: $x= \alpha_1$, $y=\tilde{y}$. From the equation({\ref{a1=a2}}), (\ref{alpha1}), (\ref{alpha2}), We know that no points belongs to class $C_1^{-}$ or class $C_2^{+}$ and all of them belongs either to class $C_1^{+}$ or class $C_2^{-}$. Thus, according to definition 3,

\begin{equation}
    g(c) = \max\{{\mbox P}(~C_1^{+})+{\mbox P}(~C_2^{-}),~{\mbox P}(~C_1^{-})+{\mbox P}(~C_2^{+})\}
    = \max\{1, 0\}
    = 1
\end{equation}

\end{proof}

\noindent All correlation coefficients described in this paper are invariant
to linear transformations of the form $\,w=av+d\ (a>0,~d\in\mathbb{R})\,$. For the Pearson correlation coefficient the proof is given
in \cite{rohatgi1976introduction} and for the other correlation measures, the proofs are straight forward and hence omitted.

\bigskip

\subsection{Estimation of $g$-Correlation}\label{est g}

For a given $\,\{(x_i,~y_i):i~=~1,~2,~\dots,~n\}$
of measurements, the $g$-correlation coefficient $\omega$ can only
be estimated as described next. Consider dividing the data
set into two subsets: a training set $T$ of size $q$
and an evaluation set $E$ of size $\,(n-q)$.
First, estimates for the separating lines $\,y=\tilde{y}\,$
and $\,x=c\,$ with an appropriate value of $c$ is found based on the training data set $T$.
For the median $\tilde{y}$ of $Y$, the sample median
\begin{equation}
    \tilde{y}:=\left\{
    \begin{array}{cc}
y^{'}_{\frac{n+1}{2}} & \text{for}\phantom{a} n \phantom{a}\text{odd}\\
\\
\frac{y^{'}_{\frac{n}{2}}+y^{'}_{\frac{n+1}{2}}}{2} & \text{for} \phantom{a} n \phantom{a} \text{even},
    \end{array}\right.
\end{equation}
where $\,(y^{'}_1,y^{'}_2,~\dots,~y^{'}_n)$ denotes
the sequence $\,(y_1,~y_2,~\dots,~y_n)\,$ of the $y$-values
of $T$ sorted in ascending order, is used.

The following algorithm is used to compute a $c$ which gives an optimal classification for the training set $T$
of measurements with respect to the classes $C_1^{+},~  C_1^{-},~  C_2^{+},~  C_2^{-}$
defined in (\ref{4classes}) see \cite{pittner2000correlation} for an alternative method for finding
a resonably good value for $c$.

\bigskip

\noindent \underline{\bf Step 1}\, Sort all pairs in the sequence
$\,s={(x_1,~y_1),~(x_2,~y_2),~\dots,~(x_q,~y_q)}_{q \ge 1}\,$ in ascending order
based on the $x$-values.

\bigskip

\noindent \underline{\bf Step 2}\, Consider the arithmetic means of
$x$s' of all successive pairs in $s$ as possible candidates for $c$.
Start the smallest value $c$ and proceed
successively to the highest value.

\bigskip

\noindent \underline{\bf Step 3}\, For the
first candidate for $c$, count the number $p_1$ of pairs $\,(x_i,~y_i)\,$
of $s$ with $\,x_i\leq c\,$ and $\,y_i<\tilde{y}\,$ along with the
number $p_2$ of pairs with $\,x_i> c\,$ and $\,y_i>\tilde{y}$.
For all other candidates update $p_1$ and $p_2$
based on whether the pairs passed since using the previous candidate
belong to $C_1^{+}$ or $C_2^{-}$.

\bigskip

\noindent \underline{\bf Step 4}\, Store the maximum classification
percentage $\,\max\{p_1+p_2,~q-p_1-p_2\}/q\,$ achieved for the test dataset $E$ along with the
corresponding candidate for $c$. Go to Step 2.

\bigskip

Finally, $\omega$ is approximated based
on the calculated values $\tilde{y}$ and $c$ using Definition 3 for the dataset $T$.


\section{MULTIDIMENSIONAL $g$-CORRELATION}\label{section5}

The multidimensional correlation problem consists in
determining whether there exists a correlation between a random vector
$\,(X_1,~X_2,~\dots,~X_M)\,$ of independent variables and
a single dependent random variable $Y$.
From all of the correlation coefficients described in this article
only Pearson correlation coefficient \cite{wesolowsky1976multiple} and the $g$-correlation
coefficient $\omega$ can be generalized for the multidimensional situation.

When generalizing the $g$-correlation coefficient to $M$ independent
variables, the line $\,y=\tilde{y}\,$ becomes a hyperplane while
the classes $C_1$ and $C_2$ become halfspaces,respectively. In order to separate
the orthogonal projections $\,(x_1^i,~x_2^i,~\dots,~x_M^i)\,$ of a set
$\,\{(x_1^i,~x_2^i,~\dots,~x_M^i,~y^i):i~=~1,~2,~\dots,~n\}\,$ of
measurements onto the $r$-dimensional space of the independent variables,
one cannot use a line similar to $\,x=c\,$ as in equation (\ref{crit}).
Instead, a hyperplane (a plane for $\,M=2\,$ and a straight line
for $\,M=1$) is sought for separating the orthogonal projections
$\,(x_1^i,~x_2^i,~\dots,~x_M^i) (i~=~1,~2,~\dots,~n)\,$ with respect to the
classes $C_1$ and $C_2$
to which the corresponding measurements $\,(x_1^i,~x_2^i,~\dots,~x_M^i,~y^i)\,$
belong. See \cite{pittner2000correlation} for further details about the multidimensional $g$-correlation
and its practical application using Fisher linear discriminant functions. Also the multidimensional $g$-correlation
coefficient is directly related to a prediction model which allows inference of
$Y$ from the realizations of
$\,X_1,~X_2,~\dots,~X_M$.


\section{CORRELATION COEFFICIENTS COMPARISON}\label{section6}

\subsection{Comparison on linearly correlated datasets}\label{comp_linear}

As so far, we have introduced 5 correlation coefficients from literature and a new nonlinear non-parametric correlation measure method: $g$-correlation ($\omega$). We run a comparison on 12 different 2-D simulated dataset with unique features to observe the robustness of $\omega$. 

\begin{figure*}
    \centering
    \includegraphics[scale=0.8]{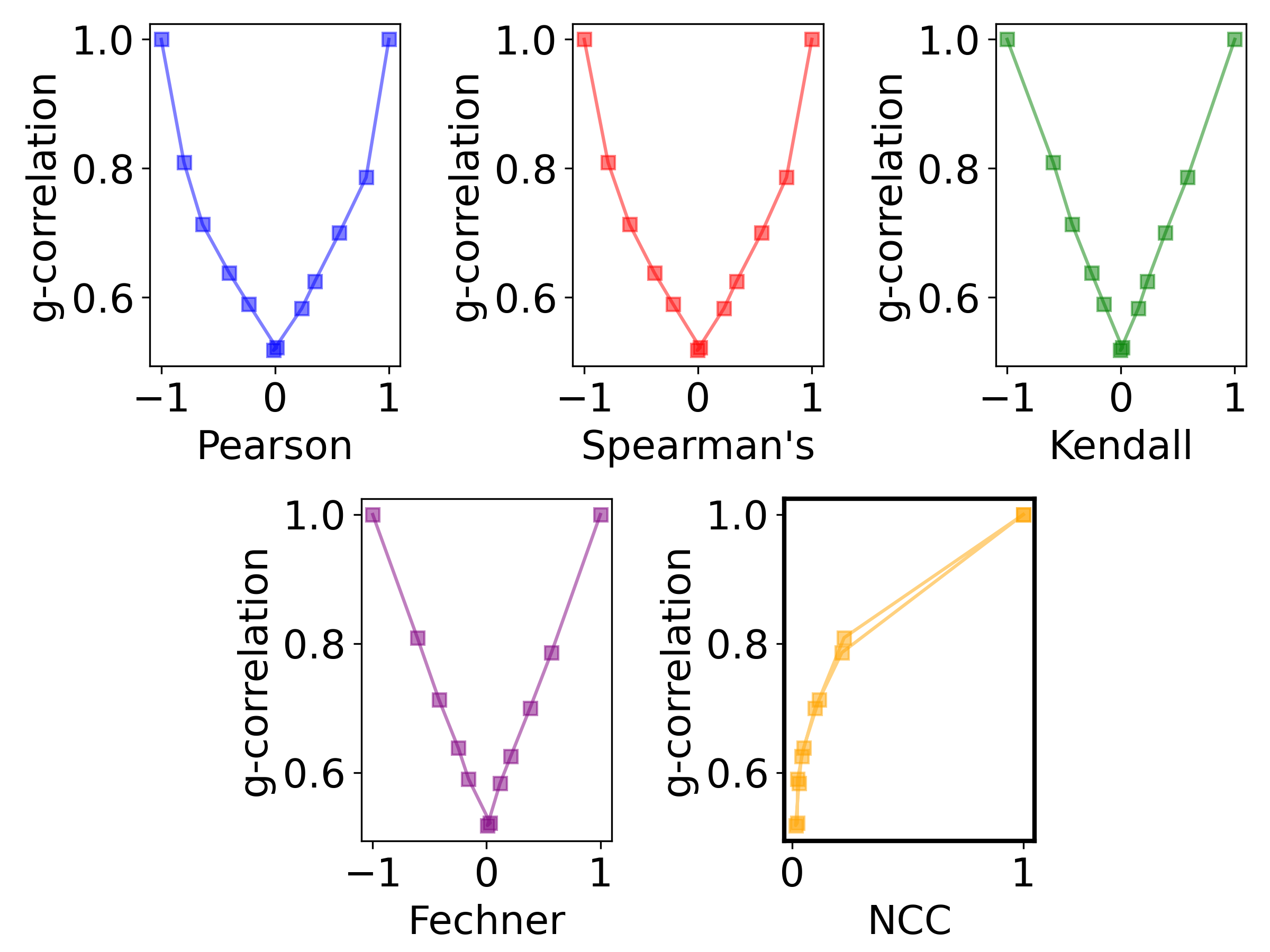}
    
    \caption{The comparison of five correlation coefficients in literature with $g$-correlation on 12 different 2-D datasets that are linearly correlated in different extend.}
    \label{fig:coeff_comp}
\end{figure*}

We can visualize the comparison from Fig.\ref{fig:coeff_comp}. The $x$-axis for each plot (from top to bottom) represents Pearson correlation coefficient ($r$), Spearman's rank correlation coefficient ($\rho$), Kendall's Tau correlation coefficient ($\tau$), Fechner correlation coefficient ($\kappa$), and Nonlinear correlation coefficient ($NCC$). The $y$-axis for all plots are $g$-correlation.

We can see that the top 4 plots share the same bowl shape that when the Pearson correlation coefficient, Spearman's rank correlation coefficient, Kendall's Tau correlation coefficient, and Fechner correlation coefficient are between [-1, 0], the $g$-correlation will decrease when these four correlation coefficient get closer to 0. This shows that $g$-correlation is robust and correct. Because when these four correlation coefficients get closer to 0, the relationship of the datasets is transforming from negative correlated to non correlated. Thus $g$-correlation is changing from its maximum, which is 1, to, its minimum, which is 0.5. 

Similarly, when the Pearson correlation coefficient, Spearman's rank correlation coefficient, Kendall's Tau correlation coefficient, and Fechner correlation coefficient are between [0, 1], the $g$-correlation will increase when these four correlation coefficient get closer to 1. Again, this proves $g$-correlation's robustness. Because when these four correlation coefficient get closer to 1, the two variables in dataset are positively correlated. Thus $g$-correlation is getting closer to 1. 

The last plot is the comparison between Nonlinear correlation coefficient and $g$-correlation. As we mentioned in section \ref{NCC}, the range for NCC is [0,1], thus . We can see from the plot that NCC and $g$-correlation are monotonic increasing. This also indicate that $g$-correlation is correct since when NCC is closer to 1, the corresponding dataset should be perfectly correlated, thus $g$-correlation should also be 1. 

Thus, in summary, $g$-correlation is robust based on the result and analysis from the experiments with 12 datasets and comparison with 5 existing correlation coefficient measurements, which vary from linear correlation coefficient to non-linear correlation coefficient, from parametric correlation coefficient to non-parametric correlation coefficient. 

\subsection{Comparison on nonlinearly correlated datasets}\label{nonlinear comp}

In section \ref{comp_linear}, we see that $g$-correlation is consistent with all the existing 5 coefficient correlation. In this section, we will show two examples which $g$-correlation outperforms than one or more other correlations in capturing the nonlinear relationship between variable $X$ and $Y$.

\begin{figure*}
  \centering
  \subfigure[]{
    \includegraphics[scale=0.5]{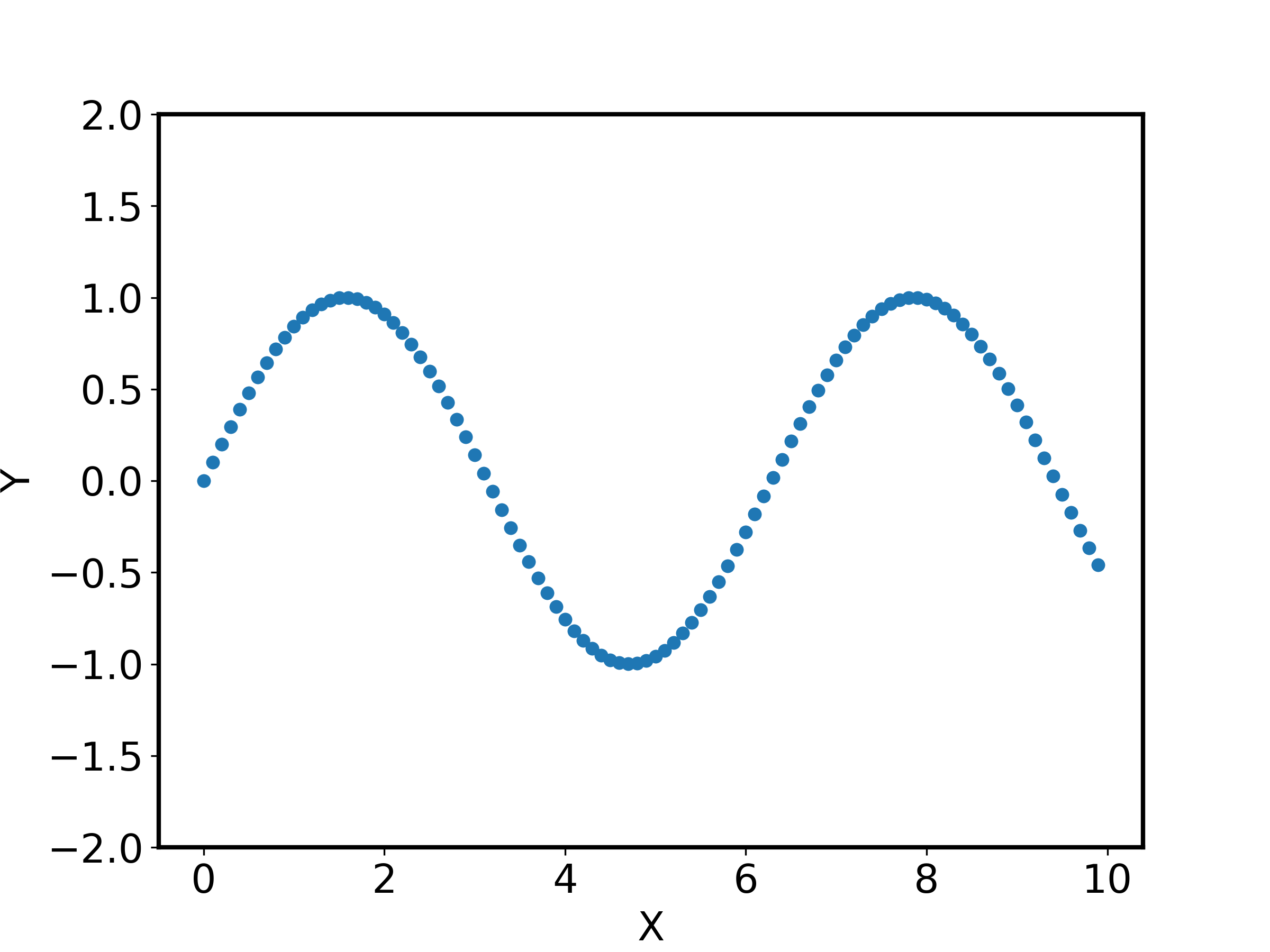}\label{fig:sin}}\\
  \subfigure[]{
    \includegraphics[scale=0.5]{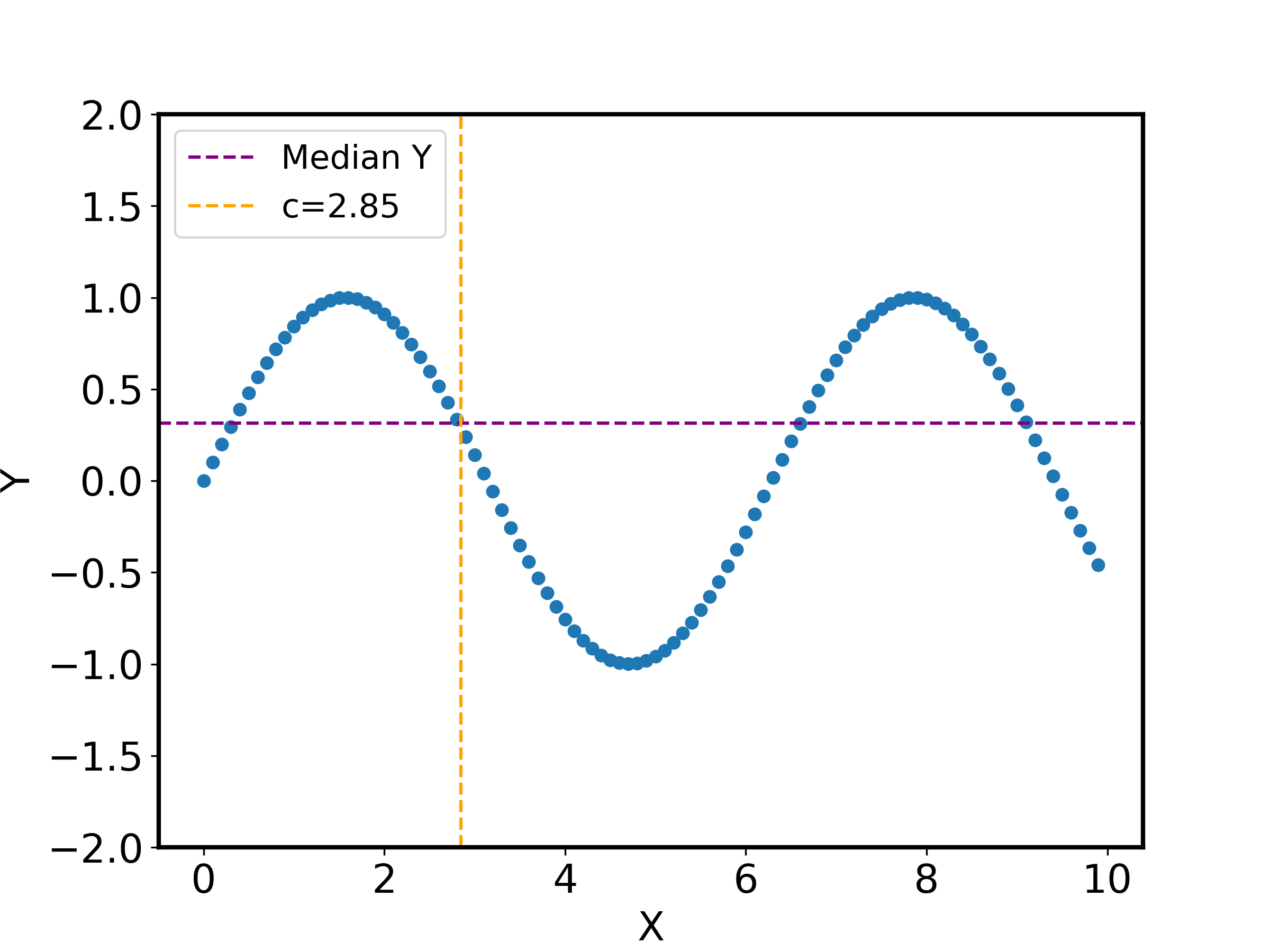}\label{fig:sin_g}}
    \caption{{(a) Nonlinear correlated random variables $X, Y$ with repeat patterns. (b) Demonstration of $g$-correlation coefficient, $g$-correlation = 0.71.}}
\end{figure*}

In the graph \ref{fig:sin}, we can see the variable $X$ and $Y$ are nonlinearly correlated and they are also auto-correlated as time series dataset, which means the pattern of the correlation repeat over certain intervals. When we exam its correlation by Pearson correlation coefficient, the result is -0.058, which is contradicted with our observation. Similarly, Spearman’s rank correlation coefficient is -0.061, Kendall’s Tau correlation coefficient is -0.042, and Fechner correlation coefficient is 0, which are all giving an inaccurate result that $X$ and $Y$ in graph \ref{fig:sin} is not correlated. 

However, $g$-correlation shows that these two variables are correlated for sure, whose coefficient is 0.71. By Lemma 4 should we know that the further $g$-correlation coefficient is from 0.5 the more nonlinearly correlated $X$ and $Y$ are. Graph \ref{fig:sin_g} demonstrates the $g$-correlation. Following the procedure in section \ref{est g}, we get the $g$-correlation by splitting the dataset with $y=\tilde{y}$ and $x=c=2.85$. 

\begin{figure*}
  \centering
  \subfigure[]{
    \includegraphics[scale=0.5]{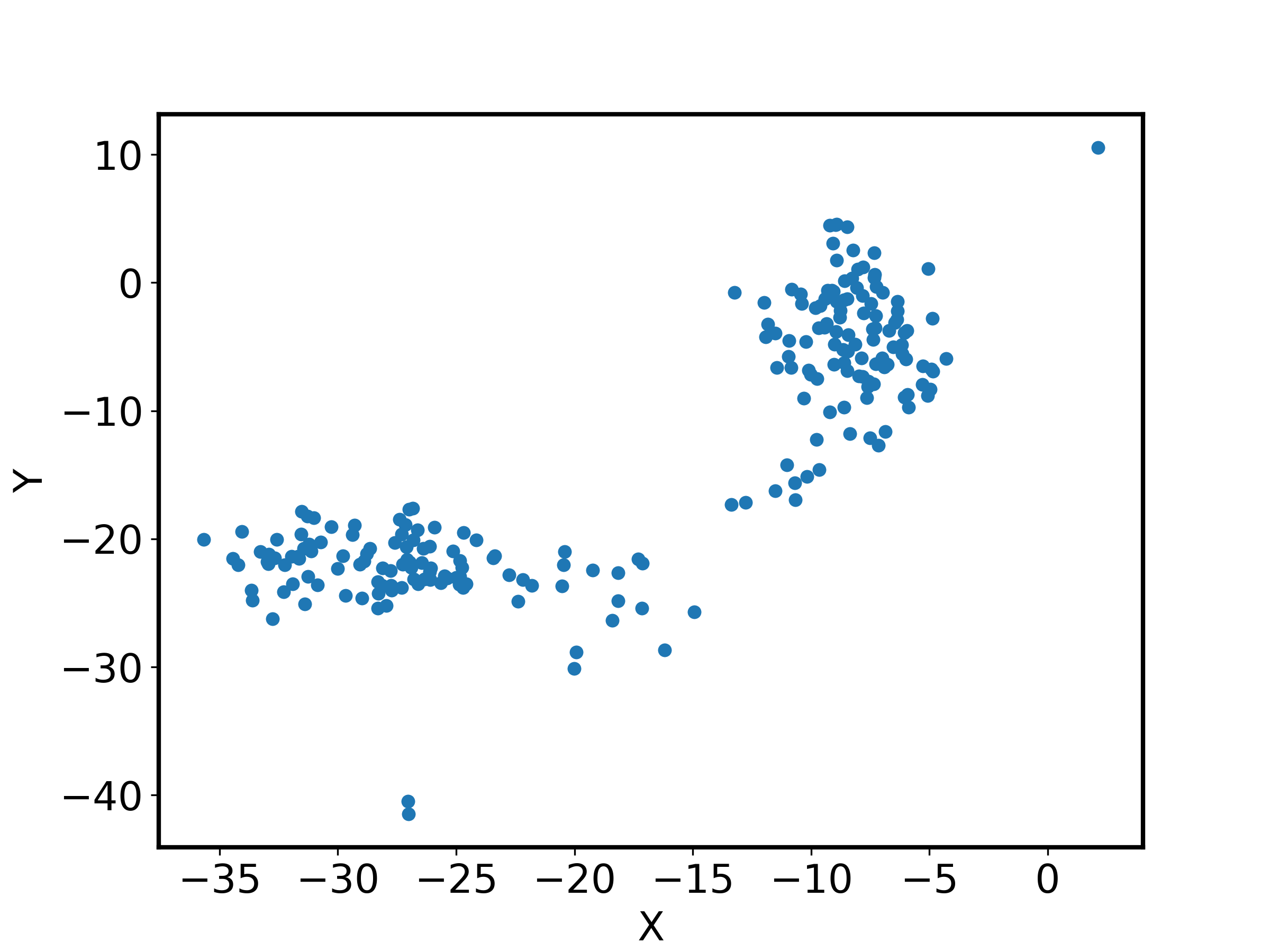}\label{fig:ncc_bad}}\\
  \subfigure[]{
    \includegraphics[scale=0.5]{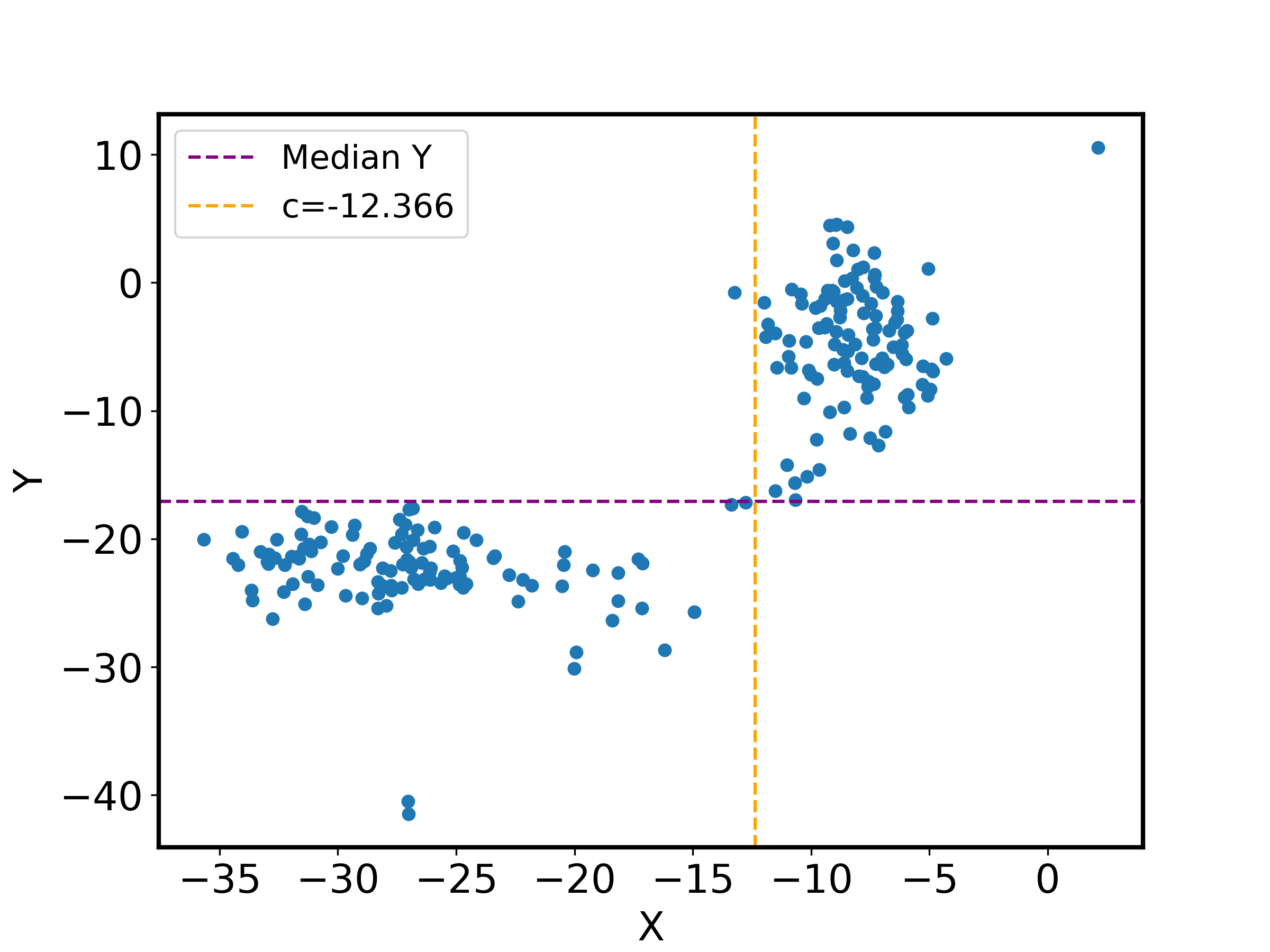}\label{fig:ncc_bad_g}}
    \caption{{(a) Nonlinear correlation example that isn't detected by $NCC$ ($NCC = 0.363$) successfully. (b) Demonstration of $g$-correlation coefficient, $g$-correlation = 1.}}
\end{figure*}

In the graph \ref{fig:ncc_bad}, we can see that variables $X$ and $Y$ has some nonlinear correlation since the range for possible $y$ varies when $x$ changes. Roughly speaking, the possible $Y$ of $x<c$ is smaller than that of $x\geq c$. By Definition 1 in section \ref{basic}, we know that this kind of dataset is correlated. And the $g$-correlation coefficient for this dataset is 1. Graph \ref{fig:ncc_bad_g} demonstrates the $g$-correlation and the optimal $c=-12.366$.

However, $(NCC)$ here is 0.363, which means that $NCC$ didn't detect the complete pattern. 

In summary, the ability for $g$-correlation captures nonlinear and complex relationship between variables is better than the 5 correlation coefficients in literature. 


\section{CORRELATION COEFFICIENTS AND SURFACE ROUGHNESS ASSESSMENT}\label{section7}

Surface roughness is an important quality indicator for products
machined with turning, milling, or grinding processes.An implementation
of adaptive control schemes requires in-process assessment of surface
roughness.
Due to the limitations of stylus profilometers, optical techniques, etc.,
surface roughness is generally measured based on the following
three parameters: arithmetic mean roughness ($R_a$), maximum peak-to-valley
roughness ($R_{max}$), and mean roughness depth ($R_z$). We use all the correlation
coefficients presented in this paper to determine the correlation
between the average level of the three surface roughness parameters
which act as the dependent variables, and cutting speed and cutting feed as
well as average values of the statistics RMS, absolute energy and
ringdown counts of acoustic emission signals which act as the
independent variables.

Data for 50 experiments with 25 different
operating conditions (varying speed and feed rates) were collected and
processed. For computing the $g$-correlation coefficient the
50 records were randomly divided 10,000 times into a training set $T$
of 30 records and an evaluation set $E$ of 20 records.
The arithmetic mean of the g-correlations for the
respective evaluation set is taken as $\omega$.

\begin{figure}\label{fig11}
    \centering
    \includegraphics[scale=0.8]{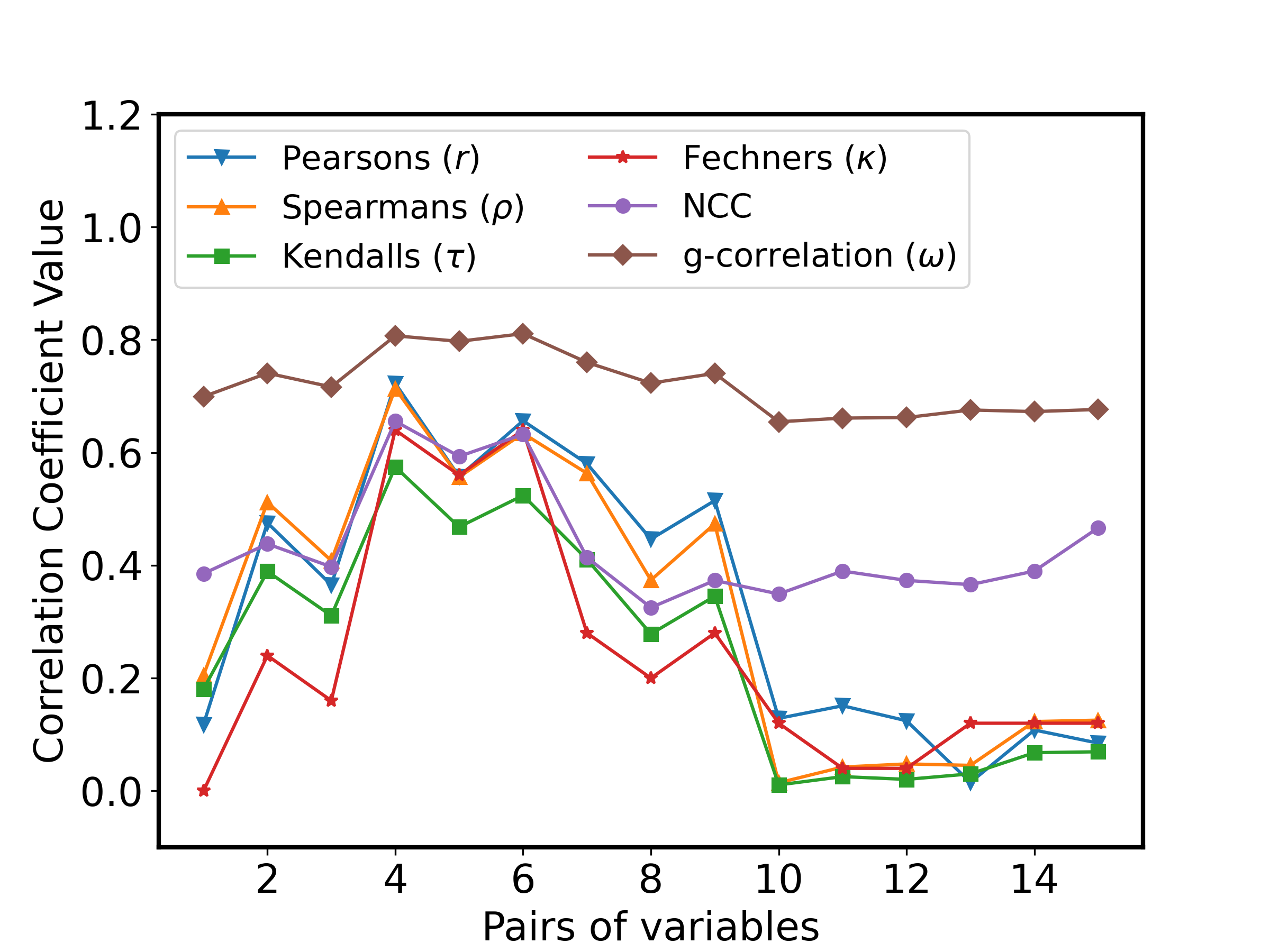}\caption{The absolute values of
correlation coefficients for surface roughness independent variables and dependent variables.}
\end{figure}

Table \ref{table1} presents the correlation coefficients for the above data sets. For identical measurements,we took the average ranks for
these equal values in finding the Spearman's rank
correlation coefficient $\rho$ \cite{rohatgi1976introduction}. Figure \ref{fig11} shows the graphical representation
of the results. Each color represents one correlation coefficient measure method. For each line, each marker is an absolute value of a correlation coefficient between one independent variable and one dependent variable. From left to right, each marker represents the correlation coefficient between cutting speed and $R_{a}$, cutting speed and $R_{max}$, cutting speed and $R_z$, cutting feed and $R_{a}$, cutting feed and $R_{max}$, cutting feed and $R_z$, one of the acoustic emission statistics RMS and $R_{a}$, one of the acoustic emission statistics RMS and $R_{max}$, one of the acoustic emission statistics RMS and $R_z$, absolute energy and $R_{a}$, absolute energy and $R_{max}$, absolute energy and $R_z$, ringdown counts and $R_{a}$, ringdown counts and $R_{max}$, ringdown counts and $R_z$,

From figure \ref{fig11} it is seen that $\omega$ has the same pattern as $\,|r|$, $\,|\rho|$, $\,|\tau|$, $\,|\kappa|$, and $NCC$. This result is consistent with the result we got from section \ref{comp_linear} with simulated dataset. In addition, in \ref{comp_linear}, we use full dataset to calculate the $g$-correlation and the result shows that the $\omega$ is consistent with $\,r$, $\,\rho$, $\,\tau$, $\,\kappa$, and $NCC$. In this section, we use real world dataset to calculate $g$-correlation by estimating the parameters $\tilde{y}$ and $c$ in training set and validating them in test set. And the result also shows that $\omega$ is consistent with $\,r$, $\,\rho$, $\,\tau$, $\,\kappa$, $NCC$. It further indicates that $g$-correlation is robust. 

However, in section \ref{nonlinear comp}, we showed that $g$-correlation outperforms when there are some complicated nonlinear relationship between independent and dependent variables. In figure \ref{fig11}, we can see that the correlation coefficients between absolute energy and $R_{a}$, $R_{max}$, and $R_z$, as well as ringdown counts of
emission signals and $R_{a}$, $R_{max}$, and $R_z$ are close to 0 based on all the correlation coefficient measurements, except for $NCC$ and $\omega$. This could be case that the hidden nonlinear relationship is captured by $NCC$ and $\omega$. 

From the standpoint of surface roughness prediction in finish turning, the
results imply that cutting feed is strongly while the cutting speed and RMS of acoustic emission
signals are moderately correlated with the three roughness parameters.
The absolute energy and ringdown counts has nonlinear correlation to surface roughness.


\section{CONCLUSIONS}\label{section8}

Several correlation coefficients have been examined in this paper, with regard to linearly and nonlinearly correlated dataset. We showed that when dealing with linearly correlated variables, $g$-correlation coefficient $\omega$ is consistent with Pearson's $r$, Spearman's $\rho$, Kendall's $\tau$, Nonlinear Correlation's $NCC$ as well as Fechner's $\kappa$. When examining more complicated nonlinear relationship, $\omega$ outperforms than all the other 5 measurements.

We also examined these correlation coefficients with regard to a problem of surface roughness assessment in finish turning. It was possible to verify former results about surface roughness prediction such as the usefulness of cutting feed through the whole spectrum of correlation coefficients. In addition, $g$-correlation is consistent with other correlation measurement methods and it can also detect some complex nonlinear relationship that most of other methods can't do.

In addition, properties of the $g$-correlation coefficient $\omega$ have been proven and an algorithm for the computation of $\omega$ has been provided.

What's more, there is no assumptions on the application of $\omega$, which makes it a universal correlation coefficient measurement method to capture either linear or nonlinear relationship. This together with the facts that it works beyond functional relationships (no parameter needs to estimate) between the data allows the $g$-correlation coefficient $\omega$ to be applied in a wide range
of areas.\\

\newpage
\begin{appendices}
\section{APPENDIX}
\begin{tabular}{@{}ll@{}} 
\textbf{Symbol} & \textbf{Description} \\ 

$X, Y$ & Random variables \\

$r$ & Pearson correlation coefficient \\

$\tau$ & Kendall's Tau correlation coefficient\\

$\rho$ & Spearman's rank correlation coefficient \\

$NCC$ & Nonlinear correlation coefficient \\

$\kappa$ & Fechner correlation coefficient \\

$\omega$ & $g$-correlation correlation coefficient \\

$|x|$ & Absolute value of $x$ \\

$\bar{x}$ & Mean value of variable $X$ \\

$\tilde{x}$ & Median value of variable $X$ \\

$\simeq$ & Approximately equal to \\

$\epsilon$ & an arbitrarily very small real number\\

sup & supremum (least upper bound)\\
inf & infimum (greatest lower bound)\\ 

$:=$ & is defined to be equal to\\

$F_{X}, F_{Y}$ & Cumulative distribution function\\

$P\{X<b\}$ & Probability that $X$ is strictly less than $b$\\

$P\{X|Y\}$ & Probability of the event $X$ conditional on the event $Y$\\

$\sum_{i=1}^{n} a_i$ & $a_1+ a_2 + a_3 + \cdot \cdot \cdot + a_n$\\

$T$ & Training set\\

$E$ & Evaluation set \\

$c, \alpha_1, \alpha_2, \xi $ & Some real number\\

$f(\cdot)$ & A function that avoids a dummy variable \\

$f^{-1}(X)$ & The inverse function of function $f(X)$ 

\end{tabular}
\end{appendices}

\newpage


\bibliographystyle{plain}
\bibliography{bib.bib}





\newpage

\newpage

\newpage

\newpage

\newpage

\newpage

\newpage

\newpage
\begin{sidewaystable}\label{table1}
\centerline{\bf Table 1}
\bigskip\bigskip
\centerline{\parbox{18cm}{\small{Comparison of various correlation coefficients for an independent
variable, (one of the cutting speed, cutting feed, or one of the
acoustic emission statistics RMS, absolute energy and ringdown counts)
and a dependent variable (one of the surface roughness parameters
$R_a$, $R_{max}$, and $R_z$.)}}}
\bigskip\bigskip\bigskip
\hspace*{-1cm}
\resizebox{24cm}{!}{%
\begin{tabular}{l|c|c|c|c|c|c|c|c|c|c|c|c|c|c|c|c|c|c|}
\cline{2-19}
& \multicolumn{6}{|c|}{} & \multicolumn{6}{|c|}{} & \multicolumn{6}{|c|}{}\\
& \multicolumn{6}{|c|}{$R_a$} & \multicolumn{6}{|c|}{$R_{max}$} & \multicolumn{6}{|c|}{$R_z$}\\
& \multicolumn{6}{|c|}{} & \multicolumn{6}{|c|}{} & \multicolumn{6}{|c|}{}\\
\cline{2-19}
& $r$ & $\rho$ & $\tau$ & $\kappa$ & $NCC$ & $\omega$ & $r$ & $\rho$ & $\tau$ & $\kappa$ & $NCC$ & $\omega$ & $r$ & $\rho$ & $\tau$ & $\kappa$ & $NCC$ & $\omega$\\
\hline
\multicolumn{1}{|l|}{ Speed} & $-0.118$ & $-0.205$ & $-0.180$ & -0.000 & 0.385 & 0.699 & $-0.475$ & $-0.512$ & $-0.389$ & $-0.240$ & 0.438& 0.741 & $-0.365$ & $-0.408$ & $-0.311$ & $-0.160$ & 0.397 & 0.716\\
\hline
\multicolumn{1}{|l|}{ Feed} &   0.723 &   0.713 &  0.575 &  0.640 & 0.656& 0.807 &  0.557 &  0.557 &  0.468 &  0.560 & 0.593& 0.797& 0.656 &  0.634 &  0.524 &  0.640 & 0.632& 0.811\\
\hline
\multicolumn{1}{|l|}{ RMS} & $-0.580$ & $-0.563$ & $-0.411$ & $-0.280$ & 0.414& 0.760 & $-0.447$ & $-0.374$ & $-0.278$ & $-0.200$ & 0.325 & 0.723 & $-0.515$ & $-0.474$ & $-0.345$ & $-0.280$ & 0.373&0.740\\
\hline
\multicolumn{1}{|l|}{ Energy} & $-0.129$ & $-0.014$ & $-0.011$ & $-0.120$ & 0.349 & 0.655 & $-0.151$ & $-0.042$ & $-0.025$ & $-0.040$ & 0.39 & 0.661 & $-0.124$ & $-0.048$ & $-0.02$ & -0.040 & 0.373& 0.662\\
\hline
\multicolumn{1}{|l|}{ Counts} &  0.015 &  0.045 &  0.030 &  0.120 & 0.366 & 0.676& 0.108 &  0.123 &  0.068 &  0.120 & 0.39& 0.673 &  0.085 &  0.125 &  0.069 &  0.120 & 0.467 & 0.676\\
\hline
\end{tabular}%
}
\end{sidewaystable}

\end{document}